\documentclass[prd,twocolumn,superscriptaddress,amsmath,amssymb,floatfix,nofootinbib]{revtex4-2}
\usepackage{graphicx}
\usepackage{booktabs}
\usepackage{xcolor}
\usepackage[colorlinks=true,allcolors=blue]{hyperref}
\newcommand{\um}{\,\mu\mathrm{m}}
\newcommand{\Mpch}{\,h^{-1}\mathrm{Mpc}}
\newcommand{\hMpc}{\,h\,\mathrm{Mpc}^{-1}}
\newcommand{\Msunh}{\,h^{-1}M_\odot}
\newcommand{\Halpha}{\mathrm{H}\alpha}

\newcommand{\OIII}{[\textsc{O\,iii}]}
\newcommand{\OII}{[\textsc{O\,ii}]}

\newcommand{\Mpcinv}{\,\mathrm{Mpc}^{-1}}
\newsavebox{\fitbox}
\newcommand{\fitwidth}[1]{\sbox{\fitbox}{#1}%
  \ifdim\wd\fitbox>\linewidth\resizebox{\linewidth}{!}{\usebox{\fitbox}}\else\usebox{\fitbox}\fi}

\begin{document}
\raggedbottom

\title{Cross-correlating the SPHEREx extragalactic background light with CMB lensing II:\\ theoretical predictions confronted with lightcone simulations}
\newcommand{\LAPTh}{Laboratoire d'Annecy de Physique Th\'eorique (LAPTh), CNRS/USMB, 99 Chemin de Bellevue BP110,
  Annecy F-74941, France}
\newcommand{\Sussex}{Department of Physics \& Astronomy, University of Sussex, Brighton BN1 9QH, UK \vspace{0.06in}}
\author{Azadeh Moradinezhad Dizgah}
\email{azadeh.moradinezhad@lapth.cnrs.fr}
\affiliation{\LAPTh}
\author{Zucheng Gao \vspace{0.06in}}
\email{Zucheng.Gao@sussex.ac.uk}
\affiliation{\LAPTh}
\affiliation{\Sussex}

\begin{abstract}
Cross-correlating intensity maps with the lensing of the cosmic microwave background (CMB) rejects the foregrounds
and noise that do not trace the dark matter. In the near infrared, the remaining signal combines the
integrated galaxy light (IGL), the intrahalo light (IHL) and several emission lines. Accurate modeling of all three
is essential to forecast detectability and interpret measurements. We present simulated SPHEREx intensity maps of all three on a lightcone,
painted on halos from the Hidden Valley $N$-body simulation, and the CMB-lensing convergence of the same lightcone, over $100\,{\rm deg}^2$
to $z \approx 6$. We assign each halo its lines and continuum by stellar-population synthesis, in two models whose
metallicities are fixed or follow its stellar mass. Halo-model predictions from the simulation's luminosity--halo
mass relations and measured clustering, without free parameters, match the auto-spectra of the IGL and lines, and their
cross-spectra with lensing, across the band: at $2.2\,\mu$m, to within $5\%$ above $\ell \approx 400$ and $7\%$ above $\ell = 200$,
respectively. The agreement requires nonlinear clustering: linear theory describes the cross-spectrum of a line up to
$\ell \approx 1000$--$1600$ at $2.2\,\mu$m, but fails for the IGL and the convergence at multipoles of a few hundred.
With the noise of a SPHEREx deep field and of Simons Observatory lensing, the unfiltered cross-spectra, each component taken alone, reach
signal-to-noise ratios of $9$--$21$ for the lines, $20$ for the IGL and $15$ for the IHL. A filter that removes radial modes
longer than about $600$\,Mpc keeps $62\%$ of the IGL's signal-to-noise and at most $20\%$ of the lines', with
fitted amplitudes within about $1\sigma$ of the predictions before and after filtering. The
emission model changes a channel's intensity by up to a factor of three, but the line detectability only by tens of
percent.
\end{abstract}

\maketitle

\section{Introduction}
\label{sec:intro}

The extragalactic background light (EBL) collects the emission of all sources over cosmic history. In the near
infrared, its continuum comes from the stars in galaxies, the integrated galaxy light (IGL), and from the intrahalo
light (IHL) of stars stripped from galaxies into their halos~\cite{Purcell2007,Cooray2012}. On top of this continuum
lie the emission lines of star-forming galaxies, which line-intensity mapping uses to probe astrophysics
and cosmology by tracing the history of star and galaxy formation and the cosmic large-scale structure (LSS) in three
dimensions~\cite{Kovetz2017,Bernal2022}. The fluctuations of this background therefore
record how galaxies formed their stars, how much starlight lies outside them, and how both trace the dark matter distribution. Once
the resolved galaxies are masked, the measured fluctuation power exceeds what the faint galaxies that remain can supply; the excess has been
attributed to the IHL~\cite{Cooray2012,Zemcov2014} or to high-redshift sources~\cite{Kashlinsky2018}, and its origin is
still debated.

Besides the extragalactic signal, maps of this background contain Galactic emission, zodiacal light, instrument noise
and systematic errors, which are often larger than the signal itself. Cross-correlating the maps with an independent tracer
of the LSS keeps only the emission that the two share. A contaminant that does not correlate with
the tracer adds to the variance of the cross-spectrum, but leaves its mean unchanged. With galaxies of known redshift,
the cross-correlation also measures how the emission evolves with redshift~\cite{Chiang2019,ChengChang2021,Scott2022}.
The lensing of the cosmic microwave background (CMB) offers a tracer free of galaxy bias: its convergence
$\kappa$~\cite{LewisChallinor2006} maps the projected matter itself. Its cross-correlation with the near-infrared background
therefore measures the emissivity of each extragalactic component times its bias, the factor by which its
fluctuations follow those of the matter, and so offers a test of the origin of the fluctuation
excess~\cite{Feder2025}.
In the far infrared, the cosmic infrared background, the thermal dust emission of
star-forming galaxies, is detected in cross-correlation with CMB lensing at high
significance~\cite{Planck2014CIB,Holder2013,vanEngelen2015,Cao2020}, as first predicted in Ref.~\cite{Song2003}. In the
near infrared, no such measurement exists yet.

NASA's SPHEREx satellite~\cite{Dore2014,Dore2018,Crill2020} is now surveying the near-infrared background in 102
channels from $0.75$ to $5\um$, covering the whole sky and returning most often to two deep fields of
$100\,{\rm deg}^2$ near the ecliptic poles~\cite{Bock2025,Korngut2026}. Each channel receives the IGL and the IHL from every distance along the line of
sight (LoS). The emission lines, H$\alpha$, H$\beta$, \OIII, \OII{} and, from $z \gtrsim 5$, Ly$\alpha$, enter each channel
from a single narrow slice in redshift. The lines and the continuum trace the same matter, so that the
cross-correlation of a channel with CMB lensing contains both and cannot by itself separate them.

To reach the lines, the continuum must be removed or fitted. Intensity maps are usually cleaned of spectrally smooth
emission by removing the slowly varying modes along each LoS, but the lensing kernel is so broad in distance
that the cross-correlation lives almost entirely in those modes. Several studies found it lost to this filter in
practice~\cite{Obuljen2018,Li2019,Modi2019LSS,Guandalin2022,Moodley2023}. Reference~\cite{Shen2025} showed that part of
it survives: the growth of structure and the evolution of the line emission along the lightcone move some of the
cross-correlation into the faster modes that the filter keeps. The authors forecast its detectability for several emission
lines, each taken alone, setting aside interloper lines and the extragalactic continuum. In the first paper of this series~\cite{PaperI},
hereafter Paper~I, we extend the forecast to every component of a SPHEREx channel,
fitted jointly. In those joint fits, a filter that removes the radial modes longer than about $600$\,Mpc costs each component at least $90\%$ of its signal-to-noise ratio; the
lines are therefore reached by
fitting the continuum, whose spectral shape across the channels must be modeled or calibrated with its auto-spectra.
Across line-emission models, the predicted detection of the lines varies by a factor of seven to eleven.

Forecasts in Refs.~\cite{Shen2025,PaperI} make several simplifying assumptions in modeling the signal and the
noise. They describe each component with its own empirical luminosity relation and with a halo model or linear bias,
apply the filter to modeled spectra whose variation from channel to channel carries the signal that survives,
and take the covariance to be Gaussian.

In this paper, we test these assumptions with simulated maps, in which the intensity and the CMB-lensing convergence
are built from the same $N$-body simulation. We examine to what extent a halo-model prediction reproduces the auto-spectra of the maps and their
cross-spectra with CMB lensing, which ingredients it needs, and on which scales linear theory still holds. We then apply to the maps the LoS high-pass filter
of Paper~I~\cite{PaperI} and the detectability statistic of Ref.~\cite{Shen2025}, and compare the maps with the model of Paper~I. By considering two emission
models for the lines and the IGL, we also test how this modeling affects the simulated maps and their auto-spectra and cross-spectra with CMB lensing.

We start from the Hidden Valley (HV) simulations~\cite{Modi2019}, which
provide both the halo catalogs and the matter density of one box. The suite was designed for intensity mapping of the
21\,cm line of neutral hydrogen after reionization, at $z = 2$--$6$~\cite{Modi2019}. It was then used to build CO and [C\,\textsc{ii}] intensity maps,
whose line luminosities follow from the star-formation rate of each halo through empirical relations, and to test
halo-model predictions of their power spectra~\cite{MoradinezhadDizgah2022}. Both studies used the snapshots as
separate boxes, each at one redshift. Gao et al.~\cite{Gao2026} instead built SPHEREx intensity maps on a lightcone,
stitched from the snapshots. These maps contain interloper lines from other redshifts and the IGL from
all redshifts of the lightcone, and the lines and the IGL of each halo come from one stellar-population
spectrum, set by its mass, unlike in the forecasts. Reference~\cite{Cagliari2025} then used these maps to test how
well a neural network separates the angular power spectrum of a target line from those of the interloper lines and
the continuum. Reference~\cite{Roy2026a} tested the removal of the continuum by principal components on SPHEREx-like
maps from a semi-analytic lightcone, in which every galaxy carries its own spectrum and dust.

Our maps follow the construction of Gao et al.~\cite{Gao2026}, which we extend with the IHL, a second emission model and
the CMB-lensing convergence of the same lightcone. From the spectrum of each halo, we paint the H$\alpha$, \OIII{} and \OII{}
lines and the total IGL, with no galaxy masked, and add the IHL with the model of Paper~I, which follows the analytic
model of Ref.~\cite{EuclidCollab2026}. We consider two models of the line and IGL emission of a halo, which we call the \emph{simple} and the
\emph{metallicity} models. In the simple model, all halos share the same metallicities, and each line is a fixed multiple of the halo's star-formation rate, as in
Paper~I. In the metallicity model, the metallicities of the stars and the gas follow the halo's stellar mass, so that the
luminosity of a line per unit star formation changes with halo mass, and the continuum of each halo changes with the
metallicity of its stars.
Painted on the simulated halos, the maps carry their nonlinear clustering and shot noise. They cover a $10^\circ\times10^\circ$ patch, the
size of one deep field, in the 78 SPHEREx channels that see H$\alpha$ at the redshifts of the snapshots, along a
lightcone to $z = 6.25$. We add the convergence, built along the same lightcone in the Born approximation. From the
single box, we draw twenty realizations by orienting each shell independently at random. The maps have no instrument
noise or foregrounds, so that this paper is not a SPHEREx forecast, which is the subject of Paper~I.

A test of such maps needs a prediction built from the same ingredients as the maps: the halo luminosities, the measured
halo abundance and bias, the shells and channels of the lightcone, and the windows of the estimator. We call it simply the prediction, and its
comparison with the maps has no free parameter. Agreement shows that the
calculation, from the halo model to the filter and the detectability statistic, describes the sky that the simulation
contains, though not that this sky is realistic. A disagreement points to the step of the calculation or of the
construction that is wrong. The extended halo model that Ref.~\cite{MoradinezhadDizgah2022} tested on this simulation
included nonlinear matter fluctuations and biasing, through a perturbative description of halo clustering based on the
effective field theory of LSS, and stochasticity beyond the Poisson limit. Here, instead, we take the nonlinear clustering
from the box itself, keep the Poisson shot noise of point sources, and compare angular spectra on a lightcone.

Throughout, we adopt the cosmology of the HV simulation and compute the linear and halofit matter power spectra with the
Boltzmann code CLASS~\cite{Blas2011}.

The rest of the paper is organized as follows. Sections~\ref{sec:sim} and \ref{sec:theory} describe the simulated
maps and the prediction. In Sec.~\ref{sec:sim_th}, we compare the
maps with the prediction and with linear theory, and in Sec.~\ref{sec:filtered} we test the filter and the
detectability statistic and compare the maps with the model of Paper~I. Section~\ref{sec:linemodels} presents the \mbox{second} line model, and Sec.~\ref{sec:conclusions} concludes. Appendices~\ref{app:halos}--\ref{app:pixel} cover the
halo catalog, the ingredients of the prediction, the construction defects and the pixel scale.

\section{The simulated maps}
\label{sec:sim}

In this section, we describe the simulated maps and how we measure the angular spectra. The convergence and every
intensity map share one patch, $200\times200$ pixels of $3'$, and one lightcone.

\subsection{The Hidden Valley simulation}
\label{sec:hv}

We use one configuration of the HV suite, HV10240-R~\cite{Modi2019}, which follows $10240^3$ particles of
mass $8.58\times10^{7}\Msunh$ in a periodic box of $1024\Mpch$, in a flat $\Lambda$CDM cosmology close to Planck 2015
($\Omega_m = 0.309$, $h = 0.677$, $\sigma_8 = 0.822$, $n_s = 0.968$). Twelve snapshots span $z = 0.5$--$6$ in steps of
$0.5$. The R configuration draws the amplitudes of its initial modes from the Rayleigh distribution, as appropriate for
mock observations, but all public configurations share the same initial phases~\cite{Modi2019}, so that the suite
provides only one independent realization of the initial conditions.

Two products of each snapshot enter the maps. From the friends-of-friends halo catalogs we keep the halos of at least
twenty particles, $1.72\times10^{9}\Msunh$. The matter field is a $1024^3$ mesh painted from a random $4\%$ of the particles. It is stored
with its assignment window deconvolved, a division in Fourier space that amplifies the modes near the Nyquist frequency of
the mesh; in real space, these modes show a ringing, and $36\%$ of the cells are negative. The stored field is therefore usable only
through averages over many cells: when we build the convergence, we integrate the mesh along the LoS through each slab, and never read
single cells or a few sampled planes (Sec.~\ref{sec:kappa}). The subsampling also adds a white
noise of power $P_{\rm shot} = 0.025\,h^{-3}\mathrm{Mpc}^3$, which we subtract from the convergence auto-spectrum.
It does not bias the cross-spectra with the intensity maps: the noise is the difference between the subsampled matter
field and the full one, which a random selection leaves uncorrelated with any map built from the halos.

\subsection{The lightcone}
\label{sec:lightcone}

Each snapshot contributes one slab of the box to the shell of the lightcone around its redshift, and the twelve shells tile
the LoS from $z= 0.25$ to $6.25$ without gaps and without overlap. The transverse cut is a fixed angle, $L/2 \pm
\chi\theta/2$ for a box of side $L$ and a patch of side $\theta$, as a survey sees one region of the sky at every
distance $\chi$. Each slab is as deep as its shell except at $z = 0.5$, where the shell, $1141\Mpch$ deep, exceeds the
box; that slab is capped at about 1000 cells and stretched by $14\%$ onto its shell (Sec.~\ref{sec:lim}). The layout of the slabs follows the SPHEREx simulations of Gao et al.~\cite{Gao2026}, except that they avoid the stretch: they take that slab at
its own depth and extend the lightcone to $z \approx 0.14$ with a second slab cut from the same
snapshot~\cite{Cagliari2025}, which carries the clustering and the spectra of $z = 0.5$ to redshifts where the matter power is
1.5 times as large. We keep one slab per snapshot instead, and the prediction carries the stretch exactly
(Sec.~\ref{sec:lim}).

The slabs are disjoint along the LoS, but not within the box. The lightcone is nearly five box lengths long,
so that the twelve cuts overlap one another, and one box cannot fill it with independent matter. On large scales, two shells
cut from the same volume are near-copies, which the Born sum adds coherently, so that the shared slabs raise the
convergence power at low multipoles above any prediction built on independent shells (Appendix~\ref{app:defects}).

We remove this coherence by re-orienting the box independently for every shell, which is also how
Refs.~\cite{Gao2026, Cagliari2025} construct the lightcone, from slabs that are cut, rotated and stitched.
For each snapshot, we draw one random isometry of the periodic box, composed of an axis permutation, reflections and a translation by a whole number of cells, and apply it to both the halos and the matter mesh of that snapshot.
Everything within a shell is therefore preserved exactly, in particular the pairing of each intensity map with the
convergence, and only the alignment between shells is lost (Appendix~\ref{app:defects}). As a result, the shells become independent realizations of the matter, so that a line channel that straddles a shell boundary sees two unrelated skies, an effect that the prediction includes (Sec.~\ref{sec:samecons}) but that is absent from the real sky. Each draw of the isometries gives a new realization of the whole lightcone, and drawing an ensemble of such realizations is our addition to the construction. We use twenty, a number set by the computing cost, which reduces the scatter of a single patch; at the lowest
multipoles, more draws would not help, since all realizations share the large-scale modes of the box
(Sec.~\ref{sec:estimator}).

The maps project the matter shells onto the sky as they project the halos, each part at its own distance, $\theta =
\Delta x/\chi$, so that the projection is a cone. Placing a whole slab at one distance would dilate each part of it
relative to the sky of the halos, by about $2^\circ$ at the edges of the $z = 0.5$ shell. Such a dilation barely
changes the convergence power spectrum, but it decorrelates the convergence from the intensity maps and lowers every
cross-spectrum at high multipoles by an order of magnitude or more (Appendix~\ref{app:defects}).

\subsection{The convergence map}
\label{sec:kappa}

The convergence is the matter overdensity integrated along the LoS against the lensing kernel. On a
twelve-shell lightcone the integral becomes the Born sum
\begin{equation}
  \kappa(\hat n) = \sum_i \delta^{\rm 2D}_i(\hat n) \int_{{\rm shell}\ i} W_\kappa(\chi)\,d\chi ,
  \label{eq:born}
\end{equation}
where $\delta^{\rm 2D}_i$ is the matter overdensity averaged through shell $i$ and
$W_\kappa = \tfrac{3}{2}\Omega_m (H_0/c)^2 (1+z)\,\chi\,(1-\chi/\chi_\star)$, with $\chi_\star$ the distance to the
last-scattering surface. The kernel is integrated across each shell rather than evaluated at its center; for the
thickest shell, at $z = 0.5$, the central value would be $2.5\%$ too high.

Each shell therefore needs the matter overdensity averaged through it, on the $200\times200$ pixels of the patch. We
take it from the $1024^3$ matter mesh of the snapshot. The halos, painted and divided by their bias, would be an
alternative route to the convergence, but their shot noise is tens of times that of the mesh. The simulation releases
halo catalogs and matter meshes, not the particles themselves~\cite{Modi2019}. Particles could be deposited on the
pixels of the patch directly, each at its own distance and with its own kernel weight, as we paint the halos
(Sec.~\ref{sec:lim}); with the mesh, the distance has to be discretized instead. It enters the convergence twice, in the lensing kernel and in the
mapping from an angle $\theta$ on the sky to a transverse position $\chi\theta$ in the box. We therefore cut each shell
along the LoS into thin slices, between 2 and 23 cells deep, average the mesh through each slice, which is how the
stored mesh may be read (Sec.~\ref{sec:hv}), and place each slice on the sky at its own distance, through the same
mapping from depth in the box to distance that the halos follow. The slices are thin enough that placing each one at a
single distance displaces its structures by less than a tenth of a pixel, averaged over the patch, which takes 261
slices in the nearest shell, at $z = 0.5$, and eight in the most distant one. The lensing kernel changes more slowly
with distance, and we integrate it over each quarter of a shell: the slices of a quarter are averaged into one map, the
$\delta^{\rm 2D}_i$ of Eq.~\eqref{eq:born}, whose index $i$ thus runs over the quarters of the twelve shells.

Averaging a slice gives a map on the $1\Mpch$ transverse grid of the box, which we resample onto the $3'$ pixels of
the sky: the pixel at angle $\theta$ from the center of the patch lies at the transverse position $\chi\theta$, with
$\chi$ the distance of the slice. Reading the map at those single points would fold the
structure smaller than a pixel into the map, an aliasing that raises the spectrum by up to $50\%$ near the Nyquist scale
of the pixels, while a linear interpolation between cells would damp the spectrum instead, to $0.78$ of its value in the highest
bin. We avoid both by padding the Fourier transform of the map with zeros, which gives the same band-limited field on a
lattice four times finer, reading it on a sky grid twice as fine as the pixels, and removing every mode above the
Nyquist frequency of the pixels before keeping one point in two.
When comparing the prediction with the measured spectra, we do not correct the measurements for any of these steps.
The prediction includes them instead, as three transfer factors that multiply the Limber spectrum and are properties of the construction rather than of the matter.

\begin{figure}[tb]
  \centering
  \includegraphics[width=0.95 \linewidth]{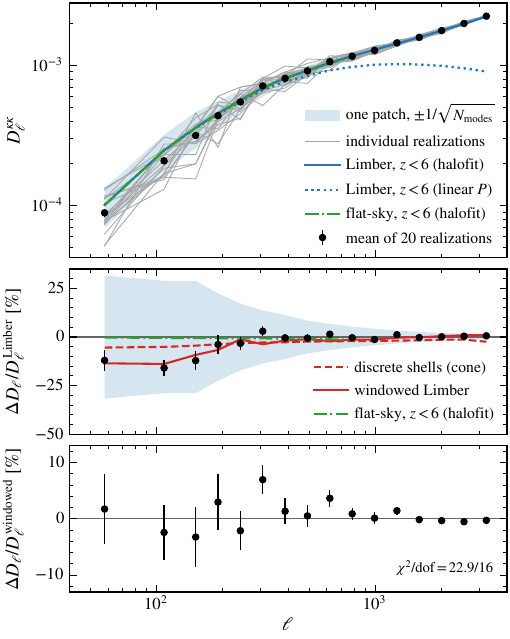}
  \caption{The convergence power spectrum. \emph{Top}: the twenty realizations (thin gray lines) and their mean
  (black points, with the error on the mean), against the Limber spectrum with the halofit (solid blue) and the linear
  (dotted blue) matter power, and the flat-sky spectrum with halofit, retaining the fluctuations along the LoS
  (dash-dotted green), all truncated at $z = 6$. \emph{Middle}: the fractional difference, in percent, from the
  baseline halofit Limber spectrum of the top panel: the mean of the realizations (black points), and two variations
  of the baseline that correct for the discreteness and the depth of the cone-projected shells (dashed red) and, in addition, for
  the transverse transfer function, giving the windowed Limber spectrum (solid red); the flat-sky spectrum is shown as
  a reference (dash-dotted green). \emph{Bottom}: the fractional difference, in percent, of the mean from the model
  prediction, the windowed Limber spectrum. The light-blue band of the top two panels shows the Gaussian
  error of one realization.}
  \label{fig:kappa}
\end{figure}

\paragraph{The discrete shells.} We reconstruct a continuous LoS with twelve shells, averaged in quarters of their depth
and each with its growth frozen at the snapshot. The averaging suppresses the radial modes shorter than the averaging depth: for a depth $\Delta\chi$, the spectrum is multiplied by $\mathrm{sinc}^2(k_\parallel\Delta\chi/2)$, with the lensing kernel applied in steps rather than continuously. Reference~\cite{Takahashi2017} derives this window and confirms it on
its own lens shells. The suppression grows as the square of the depth, so averaging in quarters rather than over whole shells cuts it
by a factor of sixteen. The residual at low multipoles is the discreteness itself, which the particles would not remove either: the
sum of twelve frozen-growth shells lies up to $1.5\%$ below the continuous Limber integral over the same redshifts. The
halofit Limber spectrum that we take as the baseline integrates from $z = 0.01$ to $6$, whereas the shells cover $z = 0.25$--$6.25$,
which lowers their spectrum by a further $3\%$ in the lowest bin and by about $1\%$ at higher multipoles.

\paragraph{The cone mapping.} Placing each part of a shell at its own distance lowers the expected spectrum by up to
$1.5\%$ below $\ell \approx 2000$. Above that, the resolution of the mesh enters: a multipole $\ell$ at distance
$\chi$ probes $k = \ell/\chi$, and the mesh holds no modes above its Nyquist wavenumber, so the nearest matter (the near edge of the lowest shell, $705\Mpch$ away) contributes nothing beyond $\ell \approx
2200$. The combined factor predicts, bin by bin, the measured change from maps with one distance per shell to maps
projected as a cone.

\paragraph{The transverse transfer function.} The remaining effects act across the sky, and each is small. The box holds only the discrete transverse modes of its lattice, and the patch is cut, interpolated, pixelized and binned. Rather than model each one, we measure them together, end to end, by passing Gaussian skies drawn on the lattice of the box through the same map-making and the same estimator. The ratio of the spectrum that comes out to the spectrum that goes in is the transfer function $T(\ell)$. It is $0.91$ in the two lowest bins, where the discrete
lattice modes matter most, and within $5\%$ of unity above them.

To validate the convergence maps, in Fig.~\ref{fig:kappa} we compare the mean of the twenty realizations, measured
with the estimator of Sec.~\ref{sec:estimator}, with the
Limber spectrum for the halofit~\cite{Smith2003,Takahashi2012} and the linear matter power, and with the flat-sky
spectrum beyond Limber~\cite{Gao2023}, drawn to $\ell = 800$, where its evaluation becomes limited by its own radial
grid rather than by the projection. The middle panel shows how the three transfer factors turn the halofit Limber
spectrum into the windowed Limber spectrum, the reference of the comparison, and the bottom panel shows the
difference of the measurement from that reference.

The simulated maps carry the full nonlinear clustering of dark matter. At low multipoles, the linear and halofit predictions
coincide, while above $\ell \approx 600$ the linear prediction flattens and the measurement follows halofit, a factor
of $2.5$ above linear in the highest bin, at $\ell \approx 3200$. That onset is where the shells that dominate the
lensing kernel reach $k \approx 0.1$--$0.3\hMpc$, the scale at which the nonlinear growth of structure begins to add
power; in the highest bin they are probed at $k \approx 1\hMpc$.

The departure of the maps from the Limber spectrum comes from the construction, not from the projection: at low
multipoles, where the Limber approximation could fail, the spectrum beyond Limber stays within a percent of it, on
the zero line of the middle panel. Against the Limber spectrum, the mean lies $12$--$16\%$ low in the three lowest
bins and within $4\%$ above $\ell \approx 200$, which the three transfer factors account for. Against the windowed
Limber spectrum, it agrees in every bin, within $7\%$, with $\chi^2 = 23$ for the sixteen bins and no bin beyond
$3\sigma$.

\begin{figure*}
  \centering
  \includegraphics[width=0.7 \linewidth]{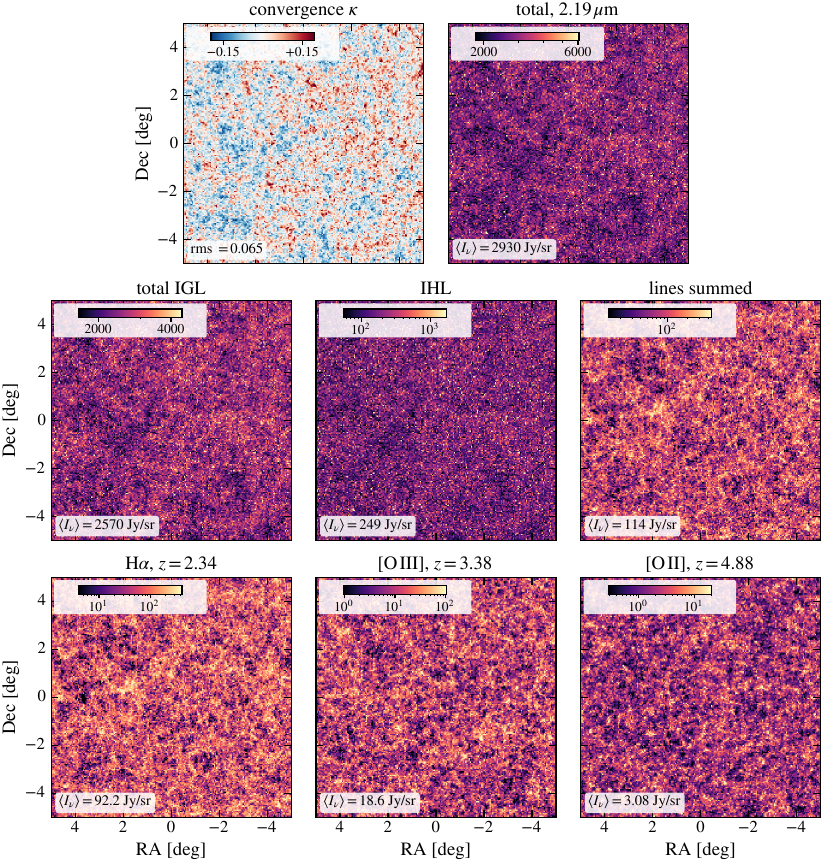}
  \caption{Comparison of the simulated CMB-lensing convergence with the EBL components, in the SPHEREx channel
  covering $2.167$--$2.219\um$, in one realization: $10^\circ\times10^\circ$ in $3'$ pixels. \emph{Top}: the
  convergence and the total EBL intensity. \emph{Middle}: the total IGL, the IHL, and the three lines summed.
  \emph{Bottom}: the three lines, each from its own redshift, H$\alpha$ from $z = 2.30$--$2.38$, \OIII\ from
  $3.33$--$3.43$ and \OII\ from $4.81$--$4.95$. The mean intensity of each EBL component and the rms of the
  convergence are given on the maps. The color bars show the brightness of each map in Jy/sr, on its own logarithmic
  scale between the 1st and 99.5th percentiles of its pixels; the convergence is on a linear scale. No smoothing is
  applied, and the maps have no point-spread function, instrument noise or foregrounds.}
  \label{fig:maps}
\end{figure*}

\subsection{The intensity maps}
\label{sec:lim}

Each halo on the lightcone carries a spectrum set by its mass and the redshift of its snapshot. We use the mean star-formation
history of halos of that mass in the UniverseMachine~\cite{Behroozi2019}, and pass it to the stellar-population code
FSPS (Flexible Stellar Population Synthesis)~\cite{Conroy2009,ConroyGunn2010} with nebular emission obtained from \textsc{cloudy}~\cite{Byler2017,Ferland2013}. The
stellar populations have a Kroupa initial mass function~\cite{Kroupa2001} and a Milky-Way attenuation
curve~\cite{Cardelli1989}, with a $V$-band optical depth of $0.2$ on all stars. From each spectrum, we take three
emission lines, H$\alpha$, \OIII\ and \OII, each integrated over its line window, and the continuum. The
latter is the IGL of Sec.~\ref{sec:intro}, built from the star-formation history of every halo rather than from a
luminosity function and with no galaxy masked, so that the maps carry the total IGL, whereas Paper~I models the
unresolved IGL left after bright galaxies are masked. A fourth
component, the IHL, is painted from the luminosity per halo of Paper~I's
model~\cite{PaperI}, which follows Refs.~\cite{Cooray2012,EuclidCollab2026}, and so shares its calibration with that
halo model.

Every halo is painted as a point into the pixel that contains it, the IHL included. A line falls in the channel that
its redshift selects, while the IGL and the IHL fall in every channel, each read from the halo's spectrum at the
rest wavelength that the channel sees at the halo's distance. All components share one log-normal scatter draw per
halo, of width $\sigma = 0.3$\,dex. The maps are in Jy/sr, on the 78 SPHEREx channels that cover H$\alpha$ at $z =
0.5$--$6$ ($1.01$--$4.56\um$); in each channel, \OIII\ and \OII\ fall from other, higher redshifts. The $3'$ pixels,
three times coarser than those of Ref.~\cite{Cagliari2025}, carry multipoles up to $\ell \approx 3600$, which suffices
for a cross-correlation with CMB lensing, as we show in Appendix~\ref{app:pixel}. The patch holds $2.5\times10^9$
halos over the whole lightcone, between $0.7\times10^8$ and $3\times10^8$ per shell, so that each pixel sums some $6\times10^4$
of them along its LoS.

Unless stated otherwise, the maps are painted with the simple model. Its three lines are each a fixed
fraction of one nebular spectrum, at a fixed gas metallicity and ionization parameter, so that their luminosities are
proportional functions of halo mass, with $\OII/\Halpha = 0.28$ and $\OIII/\Halpha = 0.53$ to a few percent over the
masses that carry the light. The stars sit at the lowest metallicity of FSPS, $Z = 4.5\times10^{-5}$, while the
nebular emission uses the solar slice of the photoionization grid; these are the defaults of the code, as in the
construction of Ref.~\cite{Gao2026} that we follow. The ratio of each line to the current star-formation rate (SFR) of the halo's
mean history is $L/{\rm SFR} = (5.9,\,3.1,\,1.7)\times10^{41}\,{\rm erg\,s^{-1}}$ per $M_\odot\,{\rm yr^{-1}}$ for
H$\alpha$, \OIII{} and \OII, values set by the metallicity of the stars.

Unlike in the CO and [C\,\textsc{ii}] maps painted on the HV simulation~\cite{MoradinezhadDizgah2022}, our halos carry a
spectrum rather than a luminosity from an empirical relation, so that the maps also hold the total IGL, fed by
the whole star-formation history. Those maps carried a log-normal scatter of $0.37$\,dex between halos, close to our
$0.3$\,dex. The maps of Gao et al.~\cite{Gao2026} had none at fixed mass and redshift, and Ref.~\cite{Cagliari2025}
applied one to the intensity of each pixel instead. The level at which the scatter acts changes the spectra: a draw
per halo is a property of the source, raising the shot noise relative to the clustering and correlating the
components of a halo, whereas a draw per pixel adds a white term whose size depends on the pixel size.

One shell is painted from a slab shorter than itself. The $z = 0.5$ shell is deeper than the box, so its slab is
stretched by $14\%$ onto it (Sec.~\ref{sec:lightcone}) and its halos lie at $f = 0.876$ of the density of the box.
Stretching a field thins it and dilates its structures at once, so that the mean intensity and both terms of the
auto-spectrum of that shell carry a factor $f$, while its cross-spectrum with $\kappa$, whose matter is stretched by the same mapping, carries none. The six bluest $\Halpha$ channels lie inside the slab and test these factors
(Sec.~\ref{sec:bands}).

Figure~\ref{fig:maps} shows the components of one channel, $2.167$--$2.219\um$, in one realization. The three lines
reach it from three redshifts, so each line map is the cosmic web of a single thin slice in redshift, while the IGL, the
IHL and the convergence integrate over every distance along the lightcone. The IGL carries $88\%$ of the mean intensity of the channel, the IHL $8\%$ and the three lines $4\%$,
at $2570$, $249$ and $114$\,Jy/sr. At $3'$ the spread between pixels is mostly due to the discreteness of the sources rather than to their clustering: the
IGL varies by a factor of a few, the IHL by more than a decade and the lines by two decades.

The single-channel figures use the $2.2\um$ channel. It holds all three lines, H$\alpha$ from $z = 2.34$, \OIII\
from $3.38$ and \OII\ from $4.88$, and is the channel in which Paper~I compares its model with Ref.~\cite{EuclidCollab2026} in the $K$ band; no line is brightest there (Sec.~\ref{sec:bands}).

We note two features of the construction that limit what these comparisons can say about the real sky. First, all twenty realizations re-orient
one $1024\Mpch$ box, so they share its large-scale modes (Sec.~\ref{sec:estimator}) and reuse its volume in shells
that only the isometries make independent (Sec.~\ref{sec:lightcone}). Second, as in the construction we
follow~\cite{Gao2026,Cagliari2025}, the maps carry no beam, point-spread function, noise, zodiacal light or Galactic
emission, so the detectability statistic of Sec.~\ref{sec:filtered} adds the noise of each channel and of the lensing
reconstruction from theory (Sec.~\ref{sec:shen}). The convergence likewise stops at $z = 6.25$, where the simulated
matter ends, and the prediction is truncated with it.

\subsection{The estimator and its error bars}
\label{sec:estimator}

We measure the angular spectra with the flat-sky estimator: the two-dimensional Fourier transform of the patch,
averaged in rings of constant $|\boldsymbol\ell|$, with $D_\ell = \ell(\ell+1)C_\ell/2\pi$ quoted throughout. The
pseudo-$C_\ell$ estimators of survey analyses correct for a mask and for the curvature of the sky; a simulated flat
patch needs neither. The multipole bins are logarithmic, merged where needed so that each holds at least ten
Fourier modes, which gives sixteen bins from $\ell = 32$ to $3600$ and a Gaussian scatter per bin of at most
$\approx 30\%$ for an auto-spectrum.

On a field that is periodic across the patch, the estimator is exact; our patch is instead cut from a larger sky, and
the cut leaks power between bins. The leakage shifts the mean bandpower, by a transfer factor that for the auto-spectra of the $2.2\um$ channel is $0.93$--$0.97$ at $\ell = 58$ and within $1\%$ of unity above $\ell
\approx 400$, and which the prediction includes. It also widens the distribution of a single patch in the lowest bins,
which hold only ten to twelve modes each: at $\ell = 58$, for a spectrum as red as the IGL's, the $95\%$ range
of the bandpower spans $0.4$ to $1.6$ times its expectation.

We estimate the error of a bandpower in two ways and, for the error bars of one patch in Secs.~\ref{sec:autos} and \ref{sec:crosses}, take the larger of
the two in each bin. The first is the Gaussian error, $\sigma(C_\ell) = C_\ell/\sqrt{N_{\rm modes}}$ for an
auto-spectrum, where $N_{\rm modes}$ counts each independent complex mode once, a mode and its conjugate together, a
count that Gaussian realizations of the patch confirm to a few percent in every bin. The second is the scatter of
the twenty lightcone realizations, which captures the non-Gaussian contributions: at high multipoles, a few bright
halos give the shot noise heavy tails, and the scatter there exceeds the Gaussian error.

At low multipoles, both estimates are incomplete, because the twenty realizations are not independent: they
re-orient one box and share its large-scale modes. For the convergence, the scatter between realizations in the lowest
bins, $23$--$27\%$, already falls below the Gaussian scatter of about $30\%$ expected for independent realizations, and the
error on the mean of the twenty, $5$--$6\%$, understates the true uncertainty further. Below $\ell \approx 400$, the
errors we quote are therefore lower bounds, and we judge a problematic bin by whether its deviation repeats across
realizations.

For a cross-spectrum measured on one realization, the Gaussian error is~\cite{Knox1995}
\begin{equation}
  \sigma^2\big(C_\ell^{\kappa I}\big) = \frac{\big(C_\ell^{\kappa I}\big)^2 + C_\ell^{\kappa\kappa}\,C_\ell^{II}}{2N_\ell},
  \label{eq:knox}
\end{equation}
with all three spectra measured on the same maps and $N_\ell = N_{\rm modes}$; for an auto-spectrum, it reduces to the
error above. Section~\ref{sec:crosses} checks this estimate against the scatter of the twenty realizations.

\section{Modeling the angular spectra}
\label{sec:theory}

In this section, we discuss the elements of the theoretical model which we compare against the measurements,
including the angular spectra, the halo model of the auto- and cross-power spectra and the tracer's biases.

\subsection{From three dimensions to the sky}
\label{sec:projection}

Each map is an integral along the LoS of a three-dimensional fluctuation $\delta_X$ weighted by a radial
kernel,
\begin{equation}
  X(\hat{\mathbf n}) = \int\! d\chi\; W_X(\chi)\,\delta_X(\chi\hat{\mathbf n};z).
  \label{eq:proj}
\end{equation}
For an intensity map, $\delta_X$ is the fractional fluctuation of the emitting population and $W_X = d\bar
I_X/d\chi$ the mean intensity the channel receives per unit comoving distance; for the convergence, $\delta_X$ is
the matter density contrast and $W_X = W_\kappa$ of Eq.~\eqref{eq:born}. Where both kernels are smooth over the
correlation length $\chi/\ell$, the angular spectrum of two such fields is well described by the Limber
approximation
\begin{equation}
  C_\ell^{XY} = \int\! \frac{d\chi}{\chi^2}\; W_X(\chi)\,W_Y(\chi)\;
  P_{XY}\!\left(k = \frac{\ell+\tfrac12}{\chi},\,z\right).
  \label{eq:limber}
\end{equation}
For the convergence, Fig.~\ref{fig:kappa} shows the accuracy of the Limber approximation.

A line channel is not smooth in this sense: its shell is $50$--$70\Mpch$ deep, shorter than $\chi/\ell$ below $\ell
\approx 60$--$100$, and Limber overestimates the line auto-spectra by about half at $\ell \approx 100$ and by a few percent
at $\ell = 1000$. In this case, we instead use the flat-sky approximation~\cite{Gao2023}, which keeps the
wavenumber $k_\parallel$ along the LoS and the two epochs of the pair,
\begin{equation}
\begin{split}
  C_\ell^{XY} ={}& \int\!\frac{d\bar\chi}{\bar\chi^{\,2}} \int\! d\delta\chi\;
    W_X(\chi)\,W_Y(\chi')\\
  &\times \int_{-\infty}^{\infty}\!\frac{dk_\parallel}{2\pi}\; e^{ik_\parallel\delta\chi}\;
    P_{XY}\!\left(\sqrt{k_\parallel^2 + \tilde\ell^{\,2}};\,z,z'\right),
\end{split}
\label{eq:flatsky}
\end{equation}
with $\chi = \bar\chi + \delta\chi/2$, $\chi' = \bar\chi - \delta\chi/2$ and $\tilde\ell^{\,2} =
\ell(\ell+1)/[\bar\chi^{\,2}(1-x^2)]$, $x = \delta\chi/2\bar\chi$~\cite{Gao2023}. The $\kappa$-cross-spectra stay in
Limber, which for a kernel that spans the lightcone is accurate to better than a percent~\cite{PaperI}, except in the channels that straddle a shell boundary (Appendix~\ref{app:ingredients}).

For the line-intensity maps, we neglect the impact of the velocities of the emitting galaxies, that is, the
redshift-space distortions (RSD), and work with spectra in real space. The cross-correlations with the CMB lensing
map are not affected by RSD, since the lensing kernel is smooth on the scale $\chi/\ell$ and confines the pair to
$k_\parallel \approx 0$~\cite{PaperI}, while on large scales the line auto-spectra of a real survey would be
weighted by the Kaiser factor~\cite{Kaiser1987}.

The three-dimensional spectra $P_{XY}$ can be computed perturbatively, with a bias expansion for the tracer, within
the halo model, or from a fitting function where one is available. For the matter we use
halofit~\cite{Smith2003,Takahashi2012}, which gives the convergence spectrum of Sec.~\ref{sec:kappa}, and for the
intensity maps we use the halo model of Sec.~\ref{sec:samecons}, which gives the Poisson term of a point-source
population directly. Reference~\cite{MoradinezhadDizgah2022} combined perturbation theory and the halo model for the CO and [C\,\textsc{ii}] maps
painted on these same halos, with the one-loop prediction for the halo power spectrum in the two-halo term.

\subsection{The halo model of the maps}
\label{sec:samecons}

Each halo is painted as a point source, and we model such a population with a halo
model~\cite{CooraySheth2002,Asgari2023}. With $L(M,z)$ the luminosity a halo of mass $M$ contributes to a channel,
$n(M,z) \equiv dn/dM$ the halo mass function per unit mass and $b(M,z)$ the halo bias, the mean luminosity density
is $\rho_L(z) = \int dM\,n(M,z)\,L(M,z)$ and the luminosity-weighted bias is $b_X(z) = \rho_L^{-1}\int
dM\,n(M,z)\,L(M,z)\,b(M,z)$. A point source has no internal structure, so the one-halo term of the auto-spectrum is
the Poisson term of the population,
\begin{equation}
  P_X(k,z) = b_X^2(z)\,P_{\rm lin}(k,z) + P^{\rm shot}_X(z), \label{eq:auto}
\end{equation}
\begin{equation}
  P^{\rm shot}_X(z) = \frac{1}{\rho_L^2}\int\! dM\,n(M,z)\,L^2(M,z) , \label{eq:shot}
\end{equation}
while the cross-spectrum with the matter keeps a one-halo term, since each source sits in a halo whose matter
follows the Navarro--Frenk--White (NFW) profile $u(k|M)$~\cite{Navarro1997},
\begin{equation}
\begin{split}
  P_{X\delta}(k,z) ={}& \frac{1}{\rho_L}\int\! dM\,n(M,z)\,L(M,z)\,\frac{M}{\bar\rho_m}\,u(k|M)\\
  &+ b_X(z)\,\mathcal{I}_\delta(k,z)\,P_{\rm lin}(k,z) ,
\end{split}
\label{eq:cross}
\end{equation}
with $\mathcal{I}_\delta(k,z) = \int dM\,n(M,z)\,(M/\bar\rho_m)\,b(M,z)\,u(k|M)$ the matter bias integral and
$\bar\rho_m$ the mean matter density. As in galaxy clustering, these are the spectra of the fractional fluctuation
$\delta_X$ of Eq.~\eqref{eq:proj}, in units of volume, $h^{-3}\mathrm{Mpc}^3$: the mean intensity enters the angular
spectra through the radial kernel $W_X$, and multiplying Eq.~\eqref{eq:auto} by $\bar I_X^2$ gives the form common in the line-intensity mapping literature, $(\bar I_X b_X)^2 P_{\rm lin}$ plus a shot noise in intensity
units~\cite{MoradinezhadDizgah2022}. The two-halo terms of Eqs.~\eqref{eq:auto} and \eqref{eq:cross} assume linear
bias on the linear matter power spectrum, which holds on large scales only. We replace them below by the clustering
of the luminosity-weighted halo field measured in the box, which carries the nonlinear matter power and the scale
dependence of the bias together, and we keep Eq.~\eqref{eq:shot} for the shot noise.

Two channels of one component, painted on the same halos, share a shot term $\propto \int
dM\,n(M,z)\,L_A(M,z)\,L_B(M,z)$~\cite{SchaanWhite2021}, with $L_A$ and $L_B$ the luminosities a halo contributes to the two channels, which
the channel covariance of Sec.~\ref{sec:shen} includes. Two components share such a term as well, which the total
covariance of Sec.~\ref{sec:threemodels} omits.

The maps have no satellites and no light profile within a halo, so Eqs.~\eqref{eq:auto} and \eqref{eq:cross} are exact for them but incomplete for
the sky, where one-halo power would add to the auto-spectra (Sec.~\ref{sec:linear}) and a diffuse intrahalo light
would have far less small-scale power.

The maps carry a log-normal scatter in the luminosity--halo mass relation of width $\sigma = 0.3$\,dex
(Sec.~\ref{sec:lim}). Following Ref.~\cite{MoradinezhadDizgah2022}, we write its moments as
\begin{equation}
  p_{n,\sigma} = \int_{-\infty}^{\infty}\! dx\,\frac{10^{nx}}{\sqrt{2\pi}\,\sigma}\,e^{-x^2/2\sigma^2} .
  \label{eq:scatter}
\end{equation}
The mean luminosity density and the cross-spectrum are multiplied by $p_{1,\sigma} = 1.27$, the two-halo term by
$p_{1,\sigma}^2 = 1.61$ and the shot term by $p_{2,\sigma} = 2.60$. All components share one draw per halo, so any
two of them carry the full $p_{2,\sigma}$ in the shot term of their cross-spectrum.

Beyond Eqs.~\eqref{eq:auto} and \eqref{eq:cross}, the prediction needs six further ingredients, of two kinds. In
three of them an analytic form fails on this box and a measurement takes its place: the Sheth--Tormen
(ST)~\cite{ShethTormen1999} mass function, the ST bias and the linear two-halo term. The other three describe features of the construction: the geometry of the lightcone, the painting of the lines channel by channel and the
windows of the estimator. The effect of omitting each is given below for the mass function, in
Sec.~\ref{sec:sim_th} for the two-halo term and the windows, and in Appendix~\ref{app:ingredients} for the others.

We use the halo mass function measured in each snapshot (Appendix~\ref{app:halos}), since with ST the prediction for the lines falls short by $13$--$18\%$ in the shot term and by $21$--$28\%$ in the clustering term. In place of $b_X(z)$, we use
the luminosity-weighted biases measured in the box (Sec.~\ref{sec:bias}). The two-halo terms also come from the box:
in each snapshot, we weight the halos by the luminosity of the component and measure the auto-power spectrum of
that field, with the shot noise subtracted, and its cross-power spectrum with the matter mesh. Their ratios to $b_X^2P_{\rm lin}$ and $b_XP_{\rm lin}$ say how far the emitters depart from linear bias on linear power. We multiply the two-halo term of Eq.~\eqref{eq:auto}
by the first ratio and replace both terms of Eq.~\eqref{eq:cross} by $b_XP_{\rm lin}$ times the second, up to $k = 0.6\hMpc$, and up to $2\hMpc$
for the IHL, whose light lies nearest, so that a given multipole probes it at nearly twice the wavenumber of the IGL.
At larger $k$, the auto-spectrum keeps its last measured ratio, or for the IHL its last measured clustering term, while the cross-spectrum returns to the halo-model shape, rescaled to
be continuous, since its one-halo term dominates there. Sections~\ref{sec:autos} and \ref{sec:linear} compare the maps with the prediction before and after this replacement.

The geometry of the lightcone enters twice: we integrate through each shell in redshift with the
growth and the spectrum held at its snapshot, and we give the stretched $z = 0.5$ slab the LoS density $f = 0.876$
of its halos.

The painting of the lines channel by channel has three consequences, since a line reaches a channel from a single thin shell.
First, that shell can cross the boundary between two snapshots, which the isometries have made independent
realizations of the matter (Sec.~\ref{sec:lightcone}). The channel is then painted from two snapshots, and its
prediction is the sum of two uncorrelated pieces, in each of which the emitters correlate with the convergence only
through the matter of their own snapshot. Second, a halo contributes its lines only if its own H$\alpha$ falls in
one of the 78 channels, so that the emitters of every line lie between $z = 0.52$ and $5.98$. Third, we project the line auto-spectra beyond Limber (Sec.~\ref{sec:projection}).

Every prediction also includes the windows of the estimator (Sec.~\ref{sec:sim_th}): the pixel window and its
aliases on the clustering term of an auto-spectrum, the pixel window and the Nyquist cut of the matter mesh on a
cross-spectrum, and the patch-cut transfer on both. The figures show beside the prediction a model with linear bias on
linear power, and each caption states which abundance and bias it uses.

\subsection{Halo bias}
\label{sec:bias}

\begin{figure}[tb]
  \centering
  \includegraphics{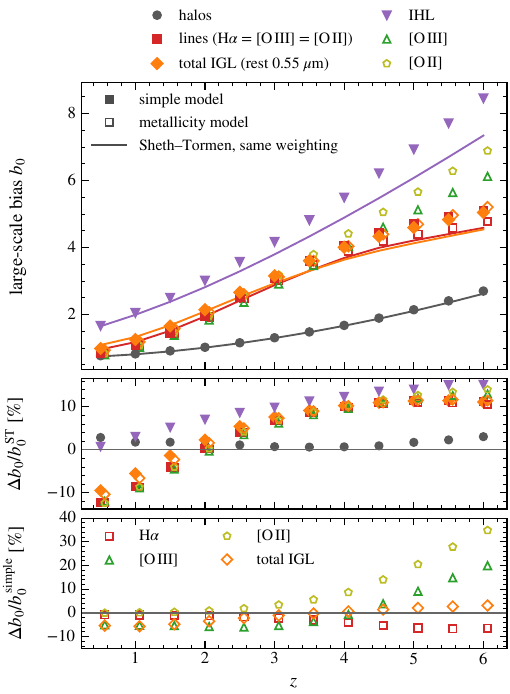}
  \caption{The large-scale bias $b_0$ of the halos weighted by each component, measured (points) and predicted by ST
  with the same weighting (lines of the same color, simple model only). Filled symbols, simple model:
  the unweighted halos (gray circles), the three lines, which share one bias (red squares), the total
  IGL at rest-frame $0.55\um$ (orange diamonds) and the IHL (purple downward triangles). Open symbols, metallicity
  model: H$\alpha$ (red squares), \OIII\ (green triangles), \OII\ (olive pentagons) and the total IGL (orange
  diamonds), shifted slightly in $z$ for legibility. Each value is fitted to the cross-spectra of the mass bins with the matter (Fig.~\ref{fig:biasmass}), summed with the luminosity of the component as weight, or with the number for the halos. \emph{Middle}: the fractional difference, in percent, of each measurement
  from its ST prediction, for both models. \emph{Bottom}: the fractional difference, in percent, of the
  metallicity model from the simple model.}
  \label{fig:biaslum}
\end{figure}

The intensity maps are halo fields weighted by luminosity, and each component weights the halos differently, so
each has its own bias. The ordering of the biases follows the masses that carry the light. Halo counts weight every
halo equally and are dominated by the smallest halos of the catalog, which are the least biased. The lines weight
the halos by star formation, which is most efficient near $10^{12}\Msunh$, so their light comes from halos well
above the catalog floor. The IGL weights them by the starlight accumulated over their history, which at low
redshift adds the massive halos whose star formation has stopped but whose old stars still shine. The IHL, counted
separately from it, weights them by the light of cluster-scale halos, and is the most biased of the four.

We measure every bias in the box, from the cross-spectrum of the weighted halo field with the matter mesh, over
all halos above the catalog floor of $1.72\times10^9\Msunh$. Since the measured bias runs with wavenumber, we
quote its large-scale limit $b_0$, from a fit $b_0 + b_2k^2$ over $k \le 0.2\hMpc$. The prediction and the linear
reference of Sec.~\ref{sec:linear} carry its average over $k < 0.3\hMpc$, on which the box ratios of
Sec.~\ref{sec:samecons} are defined; it differs from $b_0$ by at most $8\%$.
Figure~\ref{fig:biaslum} shows $b_0$ for each weighting against redshift, beside the ST prediction
for the same weight, $b_w = \int dM\,n(M,z)\,w\,b(M,z) \big/ \int dM\,n(M,z)\,w$. The luminosity weights grow more biased than the halos with redshift, since the masses that carry their light
become rarer, and more strongly clustered, as redshift increases. At
$z = 0.5$ the lines sit just above the unweighted halos and the IGL $28\%$ above them, while at $z = 6$
both are nearly twice as biased as the halos, and the IHL runs from twice the halo bias to more than three times. The IGL
and the lines approach each other with redshift, by an amount too small to read off the figure:
the IGL is the more biased of the two by $17\%$ at $z = 0.5$, where its weight relative to that of the lines
rises fivefold from $10^{11}$ to $10^{12.5}\Msunh$, by $3\%$ at $z = 3$, and above $z \approx 3.5$ it falls below
the lines, where old stars weight less massive halos than current star formation does.

The unweighted halos agree with ST at every redshift, while the luminosity weights do not. The lines and
the IGL lie about $10\%$ below the ST prediction at $z = 0.5$ and about $10\%$ above it at $z \ge 4.5$, and the
IHL lies above it at every redshift. Replacing the ST mass function by the measured one changes every ST prediction by at most $3\%$, since the
abundance enters both the numerator and the denominator of $b_w$, so the comparison tests $b(M)$ itself. The
luminosity weights follow the offset of $b(M)$ over the masses that carry their light, negative at $z = 0.5$ and
positive at high redshift, while for the halo counts the offsets at low and high mass nearly compensate
(Appendix~\ref{app:halos}). Our prediction uses the measured biases, and so is not affected.

The two line models weight the halos differently as well. In the simple model the three lines are fixed fractions of
one spectrum and share a single bias. In the metallicity model the line ratios vary with mass
(Sec.~\ref{sec:linemodels}), so each line takes its own bias: H$\alpha$ stays
below its simple-model value at every redshift, by $6\%$ at $z = 6$; \OIII{} lies $5\%$ below it up to $z = 3$ and
$20\%$ above it at $z = 6$; \OII{} equals it at $z \le 1$ and rises to $35\%$ above it at $z = 6$. Every weighting shares the
bias of the halos within a mass bin, so one measurement per bin serves every model, and weighted by the
luminosities of a model the per-bin biases reproduce the directly measured values to better than $0.25\%$ for the lines
and the IGL and to $0.5$--$1.5\%$ for the IHL.

\enlargethispage{\baselineskip}
Our measurement agrees with an earlier painting of the same catalogs. Reference~\cite{MoradinezhadDizgah2022} paints CO
and [C\,\textsc{ii}] onto them, with luminosities that also follow star formation, and finds $b_1 = 1.04$--$1.06$ for
its lines at $z = 1$, where our H$\alpha$ has $1.04$ on average over $k < 0.3\hMpc$ ($b_0 = 1.08$). At $z = 4.5$, its [C\,\textsc{ii}] has $b_1 = 3.78$, below the
$4.60$ of our H$\alpha$ ($b_0 = 4.44$), as expected: its $L \propto {\rm SFR}^{0.85}$ gives less weight to the most star-forming halos
than our H$\alpha$, which follows the rate of ionizing photons almost linearly.

\section{The spectra of the maps}
\label{sec:sim_th}

In this section, we compare the predictions of Sec.~\ref{sec:theory} with the auto-spectra of the five
components measured in the maps and with their cross-spectra with the convergence. We work first in the
$2.2\um$ channel, against the mean of the twenty realizations, then repeat the comparison in all 78 channels,
and close with the multipoles at which linear theory ceases to describe the maps.

\subsection{The auto-spectra at \texorpdfstring{$2.2\um$}{2.2 um}}
\label{sec:autos}

\begin{figure}[tb]
  \centering
  \includegraphics[width=\linewidth]{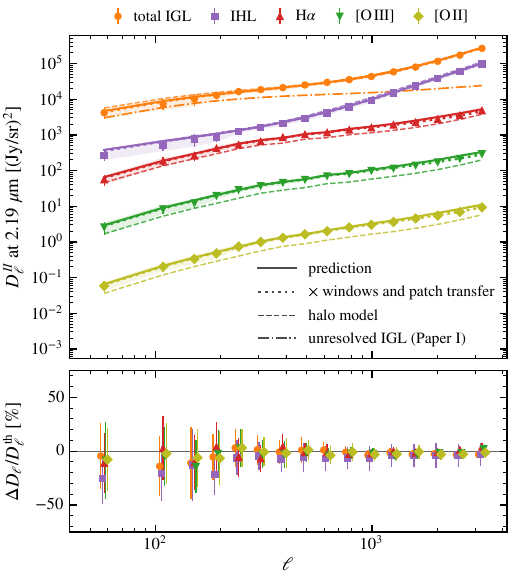}
  \caption{Auto-spectra of the five components in the $2.2\um$ channel. Points: the mean of the twenty
  realizations, with the scatter between them as a band and the error of one patch as bars. Solid: our prediction of
  Sec.~\ref{sec:theory}; dotted: the same prediction times the windows of the estimator and the patch-cut transfer, which is the reference of the lower panel; dashed: the same luminosities in a halo model with the ST abundance and bias~\cite{ShethTormen1999}, whose two-halo
  term is $b_X^2P_{\rm lin}$. Dash-dotted: the unresolved IGL of Paper~I~\cite{PaperI}, drawn for orientation only, since it
  masks galaxies brighter than $20$\,AB at $1\um$ while the maps paint every halo. \emph{Lower panel}: the fractional difference,
  in percent, of the measurement from the dotted curve, with the components offset slightly in $\ell$ for legibility.}
  \label{fig:auto22}
\end{figure}

In Fig.~\ref{fig:auto22} we show the auto-spectra of the five components in the $2.2\um$ channel. The IGL and the
three lines agree with the prediction to within $5\%$ above $\ell \approx 400$,
with a scatter between realizations of $4$--$12\%$ at $\ell \approx 500$--$800$ and $3$--$7\%$ above. Every ingredient
of the construction contributes: omitted, each would leave an offset of a few to tens of percent
(Appendix~\ref{app:ingredients}). Below $\ell \approx 400$ the lines fall short, a residual to which we return at the
end of Sec.~\ref{sec:bands}. The IHL sits $4\%$ below the prediction above $\ell = 900$, within the spread we
expect of it: its shot term is carried by the few most massive halos of each patch, each with its own log-normal
factor, so the ensemble mean moves by about that much between independent sets of scatter draws.

We reach this level of agreement only by taking the two-halo terms from the box. With linear bias on linear power
instead, and every other ingredient kept, the maps exceed the prediction by about $15\%$ in the IGL auto-spectrum over
$\ell \approx 250$--$800$, below its shot-dominated scales, and by $15$--$30\%$ in the line auto-spectra over $\ell \approx
1000$--$3000$, far beyond the scatter between realizations (Sec.~\ref{sec:linear}). To locate the failure we measure in
the box the auto-power spectrum of the IGL-weighted halo field and its cross-power spectrum with the matter. Both
exceed linear theory by nearly the same amount, about $10\%$ over $k = 0.1$--$0.25\hMpc$ and $30\%$ over
$0.25$--$0.5\hMpc$. Once the box's own matter power replaces the linear one, both return to within $7\%$ of linear bias
over the first range and $13\%$ over the second, so the painted IGL is close to a
linearly biased tracer of the nonlinear matter field. Its auto-spectrum departs from linear theory already at low
multipoles, since four-fifths of its clustering term comes from the two nearest shells, where a given multipole
corresponds to a larger wavenumber.

The lines need more than the nonlinear matter power, since the halos that carry them are biased nonlinearly as
well. At $z = 5$, where \OII{} enters the channel, its auto-spectrum exceeds the square of its bias times the matter power
of the same snapshot by $63\%$ on average over $k = 0.25$--$0.5\hMpc$, and its correlation with the matter falls to $0.87$. The
high-redshift lines are therefore the worst case, with clustering terms up to a factor of two and a half above the
linear ones in the highest bins.

The IHL weight is the most nonlinearly biased of the five, since it sits in the rarest halos: at $z = 5$ its
clustering term is about three times $b_X^2$ times the matter power, against $1.6$ times for \OII{} there. Its light
comes from the nearest shells, however, where the ratio is about a half instead at $k = 0.4\hMpc$, as halos this
massive are large and cannot sit close to one another. None of this is visible in its auto-spectrum, which is $76\%$ shot noise already at $\ell \approx 490$ and
$98\%$ at $\ell \approx 992$; the IHL departs from linear theory in its cross-spectrum with the convergence instead, through the
one-halo term (Sec.~\ref{sec:linear}).

The line auto-spectra also carry the pixel window of the estimator. Each halo is placed in the pixel that contains it, so the
clustering term carries the pixel window squared, with aliases returning up to $29\%$ of what it removes in the highest bin, while the white
shot term carries neither. On the total auto-spectrum the correction reaches $12$--$16\%$ for the lines in the two highest
bins and stays below $4\%$ for the shot-dominated IGL.

\subsection{The \texorpdfstring{$\kappa$}{kappa}-cross-spectra at \texorpdfstring{$2.2\um$}{2.2 um}}
\label{sec:crosses}

\begin{figure}[tb]
  \centering
  \includegraphics[width= \linewidth]{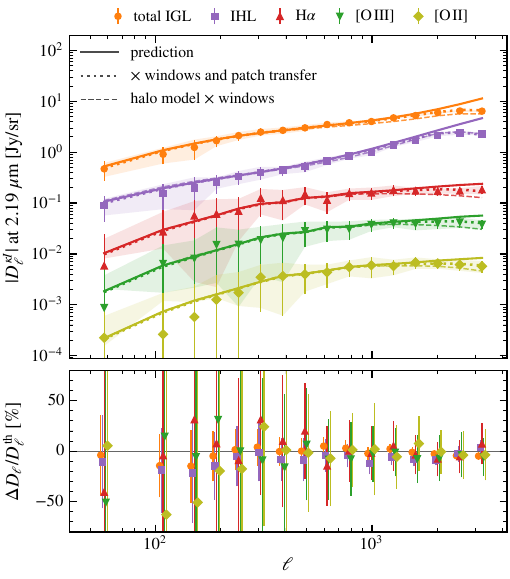}
  \caption{Cross-spectra of the five components with the convergence in the $2.2\um$ channel. Points: the absolute value
  of the mean of the twenty realizations, with the scatter between them as a band and the error of one patch as bars.
  Solid: our prediction of Sec.~\ref{sec:theory}; dotted: the same prediction times the two windows of the cross-spectrum
  estimator and the patch-cut transfer, which is the reference of the lower panel; dashed: the halo model with the measured abundance and bias, that is, the
  same prediction with linear bias on linear power in the two-halo terms, times the windows. \emph{Lower panel}: the
  fractional difference, in percent, of the measurement from the dotted curve, with the components offset slightly in
  $\ell$.}
  \label{fig:kx22}
\end{figure}

\begin{figure*}[t]
  \centering
  \includegraphics[width=0.7 \linewidth]{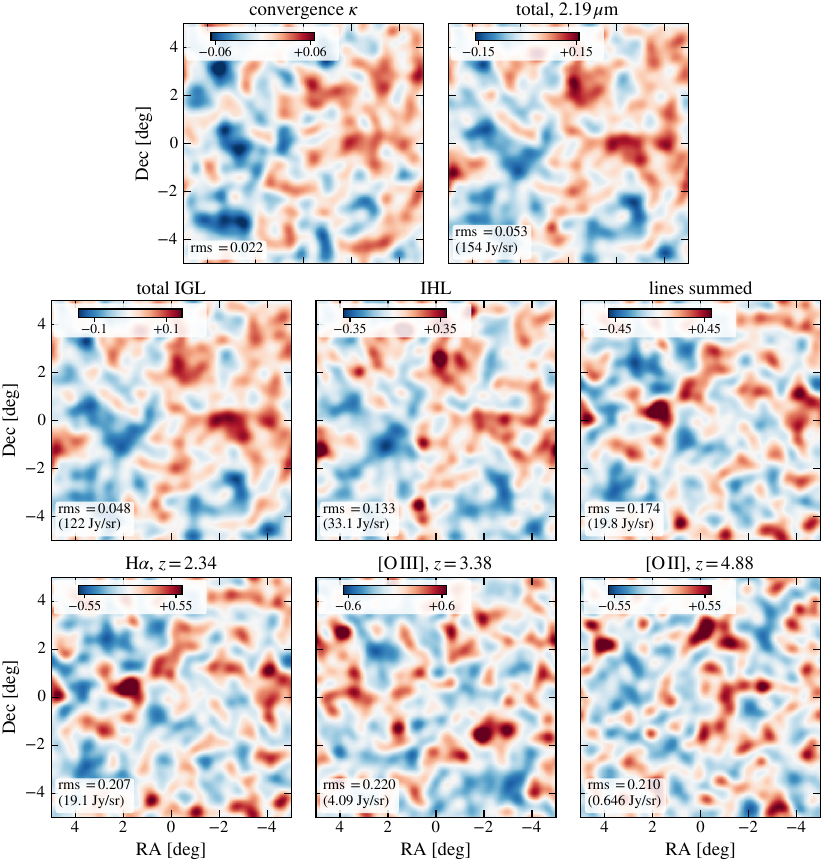}
  \caption{The maps of Fig.~\ref{fig:maps} smoothed with a $30'$ Gaussian and shown as fractional fluctuations,
  $I_\nu/\langle I_\nu\rangle - 1$, with $\kappa$ shown as $\kappa$; each map on a linear scale of $\pm3$ rms, its
  rms given as a fraction of its mean and, in parentheses, in Jy/sr. One realization.}
  \label{fig:mapssmooth}
\end{figure*}

Figure~\ref{fig:kx22} shows the cross-spectrum of each component with the convergence in the $2.2\um$ channel. Two windows of the estimator shape
the highest bins. The first is the pixel window, which against the band-limited convergence multiplies every cross-spectrum by
one power of it, down to $0.70$ in the highest bin. The second is the mesh, which holds no modes above its
Nyquist wavenumber $k_N$, so that the convergence is blind at a multipole $\ell$ to everything nearer than $\ell/k_N$. In
the highest bin that cut removes $33\%$ of the IHL's cross-spectrum and $17\%$ of the IGL's, and spares the lines, whose
shells lie far beyond.

The IGL's cross-spectrum exceeds each line's by more than an order of magnitude at every multipole, because the convergence
and an IGL map add up the same structures. The convergence sums all the matter out to the source plane, weighted by
the lensing kernel, and an IGL map at a fixed wavelength is also a sum over the whole path, since the stellar
spectrum is smooth and galaxies at every redshift contribute to the channel. A line is instead a spike in the
spectrum: H$\alpha$ reaches the $2.2\um$ channel only from a shell $68\Mpch$ deep, about $1\%$ of the path, so that
the structures in front of it and behind it, which make most of $\kappa$, are absent from the H$\alpha$ map. The IHL
is broad-band like the IGL, but is emitted mostly at low redshift, where the kernel is small.

The same ordering appears in the maps themselves, once the shot noise of the individual halos is smoothed away. In
Fig.~\ref{fig:mapssmooth} we show them smoothed with a $30'$ Gaussian and divided by their means: the IGL and the
total intensity follow the large-scale pattern of $\kappa$, the IHL less so, and the three lines hardly at all.
Each line map correlates with the convergence all the same, because the matter of its shell is itself part of what
lenses the CMB.
Over the twenty realizations the pixels of the smoothed IGL correlate with those of the smoothed convergence at $r
= 0.72$, the IHL at $0.45$ and the three lines at $0.08$--$0.14$, with a scatter of $0.03$--$0.06$ between
realizations. Dividing by the mean also removes the brightness ordering of Fig.~\ref{fig:maps}: the IGL fluctuates
the least of the five but is nearly thirty times
brighter than H$\alpha$ and so has the largest fluctuation in intensity.

With both windows, the cross-spectra agree with the prediction: the IGL to within $5\%$ at every multipole above $\ell \approx 200$ and the lines to within $7\%$ once averaged over several bins. Above $\ell \approx 1000$ their
scatter between realizations matches the Gaussian error of Eq.~\eqref{eq:knox} to better than $15\%$. The IHL is
the exception, $8\%$ short over $\ell \approx 1000$--$2000$, a residual to which we return at the end of Sec.~\ref{sec:bands}.

With linear bias on linear power in the two-halo terms, the maps would
exceed the prediction by $9$--$17\%$ for the IGL over $\ell \approx 600$--$3000$ and, for the lines, by $1$--$4\%$
at $\ell \approx 1000$, growing to $20$--$40\%$ by $\ell \approx 3000$. The auto-spectrum and the cross-spectrum of one component test the prediction
against each other as well, since the auto-spectrum carries the tracer's bias twice and the cross-spectrum once. Under linear power
the two disagree, and no single error in the bias or in the clustering power reconciles them, because the auto-spectrum and the cross-spectrum
sample the two-halo term at different scales and redshifts; with the box's clustering they agree at $\ell \approx
250$--$500$, where the test is cleanest.

The thin shell makes a line's cross-spectrum noisy as well as small. In the mean of the twenty realizations, a line
map correlates with $\kappa$ at only $r_\ell = C_\ell^{\kappa I}/(C_\ell^{\kappa\kappa}C_\ell^{II})^{1/2} = 0.04$--$0.20$, while its shot noise adds to $C_\ell^{II}$ without
adding to the signal, so that the Gaussian term of Eq.~\eqref{eq:knox} exceeds the signal term by a factor $1/r_\ell^2 \sim
25$--$600$. A detection of the cross-spectrum of a line must therefore combine many multipoles and channels.

\subsection{Across the band}
\label{sec:bands}

\begin{figure}[tb]
  \centering
  \includegraphics[width=0.95 \linewidth]{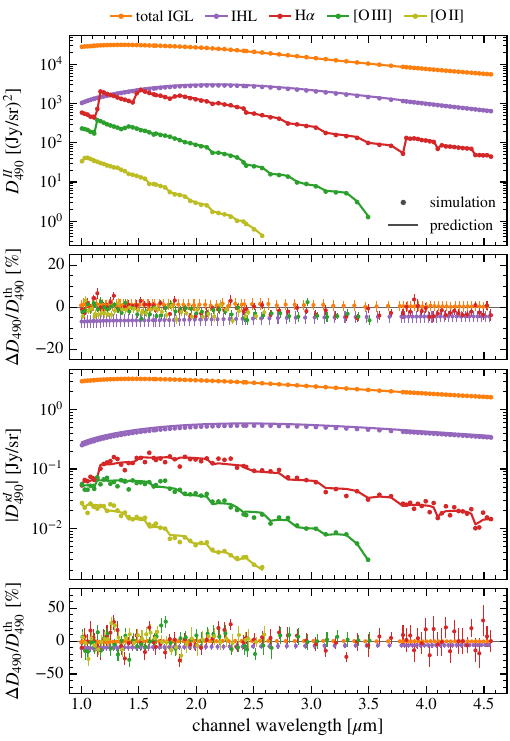}
  \caption{The five components across the 78 channels, at $\ell \approx 490$. First and third panels: their auto-spectra and
  the absolute value of their cross-spectra with $\kappa$. Points are the mean of the twenty realizations and curves the prediction times the estimator's windows; the second and fourth panels give the fractional
  difference between them, in percent, with the error on the mean. Each line's curve ends where its emitters leave
  the lightcone at $z = 5.98$, \OII{} at $2.60\um$ and \OIII{} at $3.49\um$, and steps where the line's redshift
  crosses from one snapshot's shell into the next. The H$\alpha$ auto-spectrum jumps at $3.8\um$, where SPHEREx's channels narrow
  threefold between bands.}
  \label{fig:bandprom}
\end{figure}

Figure~\ref{fig:bandprom} shows the auto-spectra and the $\kappa$-cross-spectra of the five components in all 78 channels,
in the multipole bin centered at $\ell \approx 490$. We choose this bin because the windows of the estimator are close
to unity there, and because the clustering term still dominates the auto-spectra of the IGL and the lines. Shot noise makes up only $24\%$ of
the IGL auto-spectrum and $8\%$ of the line auto-spectra. The curves of the lines have steps because the emitters of each channel come from
a single snapshot. At $3.8\um$ the SPHEREx channels become three times narrower, so a channel receives a line from a
shell three times thinner. The mean intensity of the line does not change, but its fluctuations are averaged over a
shorter path, so the H$\alpha$ auto-spectrum jumps up.

The IGL dominates every channel, in the auto-spectra and in the cross-spectra. Each line has its largest auto-spectrum in
the channels that receive it from the redshifts where its emissivity peaks: around $1.5\um$ for H$\alpha$, and at the
blue end of the band for \OIII{} and \OII{}. These maxima are broad, and even there H$\alpha$ stays more than an order
of magnitude below the IGL, and \OIII{} and \OII{} much further below. The choice of channel therefore matters for
detecting a line only once the maps are filtered (Sec.~\ref{sec:filtered}).

The prediction follows every component from channel to channel. The line auto-spectra agree with it to $2$--$3\%$ rms over
the channels, and their cross-spectra to $10$--$13\%$, about as well as the errors allow. The IGL agrees in every channel,
and the IHL lies $5$--$8\%$ low. For both, the ratio to the prediction changes by only $1$--$2\%$ across the band,
because every channel collects the light of nearly the same structures along the LoS.

Two groups of channels test single ingredients of the prediction. The six bluest H$\alpha$ channels lie inside the
stretched $z = 0.5$ slab (Sec.~\ref{sec:lim}) and agree with the prediction to within a few percent. If the
prediction treated the slab instead as the box's halos thinned at random to the same density, these channels would
lie $8$--$17\%$ above it, several standard deviations each. The reddest channels, where the emitters are the most
biased of the band, also agree to within a few percent.

We also compare each line with the prediction at every multipole, near the channel where its auto-spectrum is largest
($1.48\um$ for H$\alpha$). Each line agrees there as well as at $2.2\um$. Over $\ell \approx 1000$--$1600$, the auto-spectra
lie within $2\%$ of the prediction and the cross-spectra within $12\%$, at most two standard deviations away. In the highest
bins, the auto-spectra of H$\alpha$ and \OIII{} rise $3$--$7\%$ above the prediction, and the cross-spectra up to $16\%$.

Three residuals remain open, and none of them changes a conclusion of this paper. First, below $\ell \approx 400$ the
line auto-spectra fall a few percent short of the prediction, \OIII{} the most. Second, the IHL cross-spectrum runs $8\%$ low over
$\ell \approx 1000$--$2000$, eight times the error on the mean, and below $\ell \approx 200$ both the IHL auto-spectrum and
cross-spectrum fall further. Third, at the lowest multipoles the ratios scatter, most visibly for the IGL and the IHL, because
the cut of the patch leaks power between neighboring bins (Sec.~\ref{sec:estimator}). The twenty realizations share
the large-scale modes of one box, so they cannot tell whether this scatter hides a common deficit of $10$--$20\%$.

\begin{figure*}[htbp!]
  \centering
  \includegraphics[width=0.95 \linewidth]{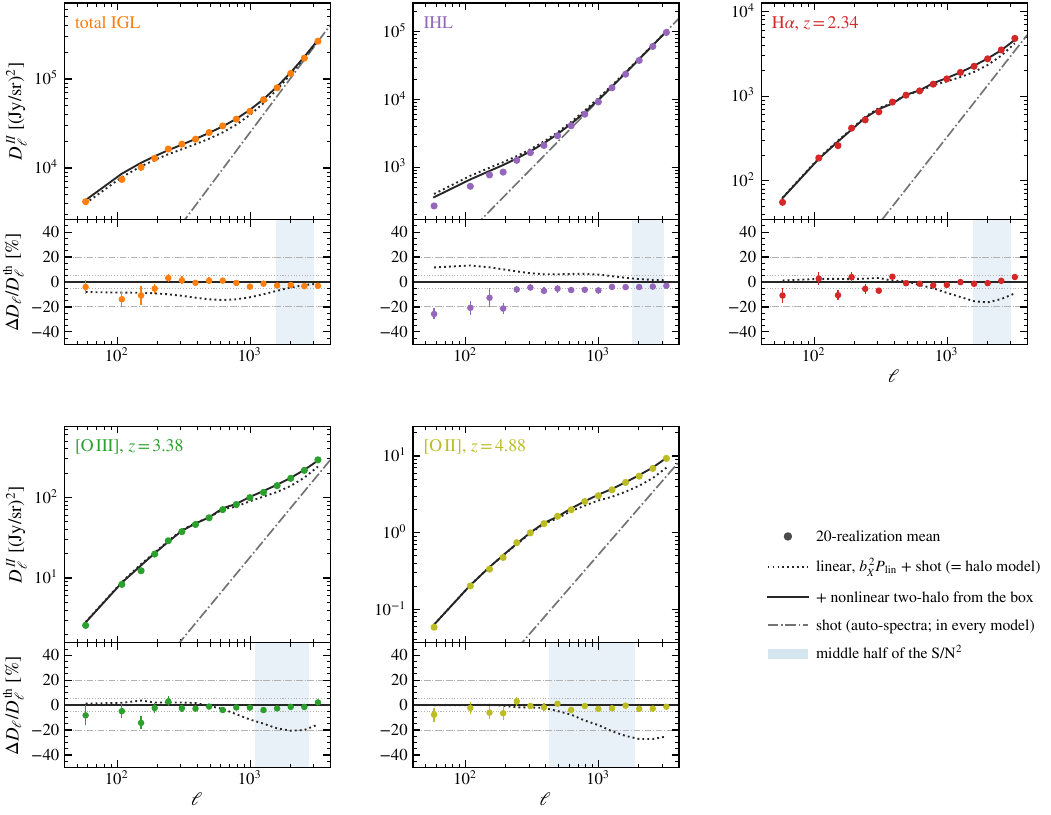}
  \caption{The auto-spectra of the maps in the $2.2\um$ channel against two predictions that share every ingredient
  of Sec.~\ref{sec:theory} and differ only in the clustering: linear theory (dotted), which for a population of
  points already carries the one-halo term as shot noise and so coincides with the halo model, and the same
  prediction with the two-halo term measured in the box (solid). Dash-dotted: the shot noise alone, which every
  model contains. Points: the mean of the twenty realizations, with the error on the mean, a lower bound below
  $\ell \approx 400$. \emph{Lower panels}: the fractional difference, in percent, from the solid prediction, for the maps
  (points) and linear theory (line), with $\pm5\%$ (dotted gray) and $\pm20\%$ (dash-dotted gray) marked. Shaded:
  the multipoles that hold the middle half of the squared signal-to-noise ratio of the auto-spectrum in one simulated deep field.}
  \label{fig:nonlinautos}
\end{figure*}

\begin{figure*}[htbp!]
  \centering
  \includegraphics[width=0.95 \linewidth]{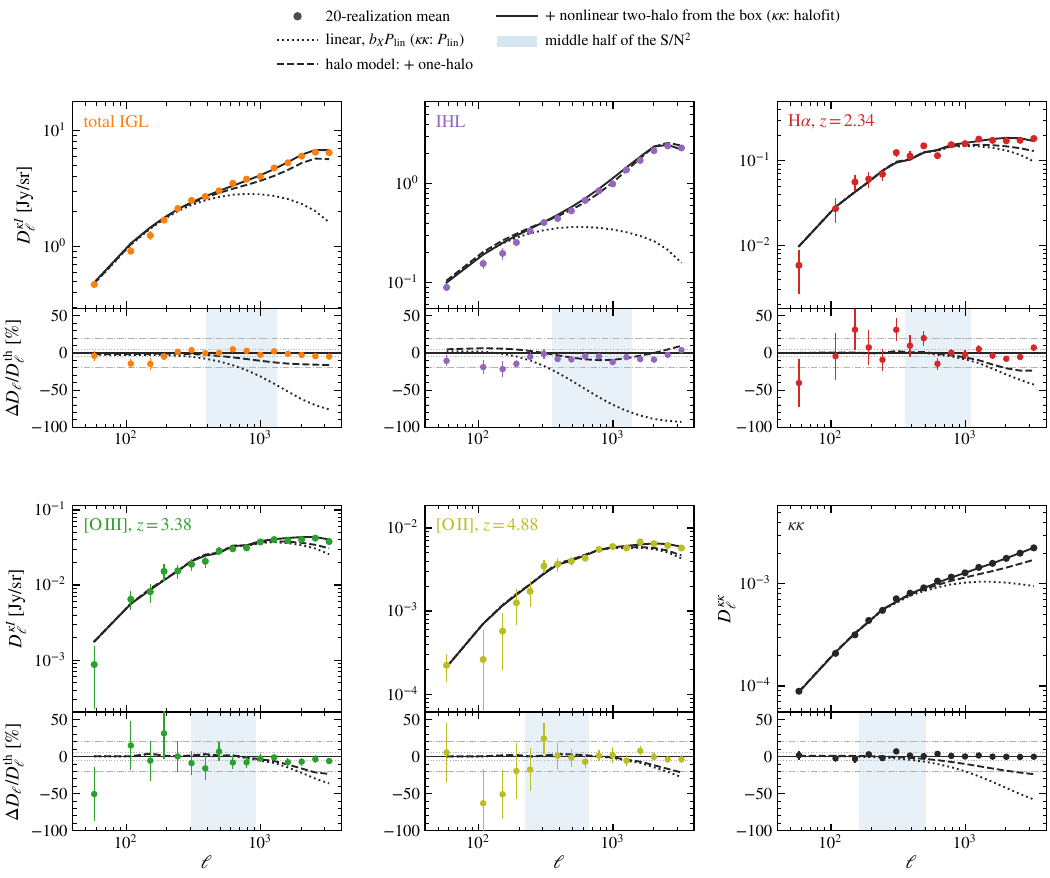}
  \caption{The $\kappa$-cross-spectra of the maps in the $2.2\um$ channel, and $\kappa\kappa$, against three predictions that
  each add one piece of physics: linear theory (dotted; $P_{\rm lin}$ for $\kappa\kappa$), the halo model with its
  one-halo term (dashed; the halo-model matter power for $\kappa\kappa$), and the halo model with the two-halo term measured in the box (solid; halofit for
  $\kappa\kappa$). Points: the mean of the twenty realizations, with the error on the mean, a lower bound below $\ell
  \approx 400$. \emph{Lower panels}: the fractional difference, in percent, from the solid prediction, for the maps (points)
  and the two simpler predictions (lines), with $\pm5\%$ (dotted gray) and $\pm20\%$ (dash-dotted gray) marked. Shaded: the multipoles that hold
  the middle half of the squared signal-to-noise ratio of the cross-spectrum, or of $\kappa\kappa$ in its own panel, in one simulated deep field.}
  \label{fig:nonlincross}
\end{figure*}

\subsection{Where linear theory fails}
\label{sec:linear}

We ask on which scales a linear description of the components holds. Line-intensity mapping forecasts, among
them Ref.~\cite{Shen2025}, treat each line as a linear tracer of the matter.
The maps contain the nonlinear clustering, the scale-dependent bias and the matter inside each halo, which that description omits.

Figures~\ref{fig:nonlinautos} and \ref{fig:nonlincross} compare the mean of the twenty realizations at $2.2\um$,
in the auto-spectra and in the cross-spectra with $\kappa$, with predictions that share every ingredient of
Sec.~\ref{sec:theory} and differ only in the clustering. The lines follow linear theory on large scales and rise above it on small ones: their auto-spectra depart from it by more than $10\%$ from
$\ell \approx 800$--$1250$ and their cross-spectra from $\ell \approx 1250$--$2000$, at $k \approx 0.2$--$0.4\hMpc$. Adding the one-halo term does not close this gap. Each
halo is painted as a point, so the one-halo term of an auto-spectrum is only its shot noise, which linear theory
already includes. A cross-spectrum with the matter also has a one-halo term, since each source correlates with the matter of
its own halo. Linear theory does not contain it, but for the lines it is small on these scales. The box's two-halo term restores agreement, so that for the lines linear theory misses the nonlinear clustering and the scale dependence of their bias, not the structure inside halos. Near their brightest channels the lines leave linear theory at lower multipoles but at the
same wavenumbers, so the limit is a physical scale.

\enlargethispage{\baselineskip}
The IGL and the IHL depart from linear theory less in the auto-spectrum than in the cross-spectrum. Their auto-spectra are
dominated on small scales by shot noise. Their cross-spectra part from linear theory already at $\ell \approx 300$--$500$ and stand well above it by
$\ell \approx 1000$, the IGL by a factor of $1.4$ and the IHL by nearly three. For them, the
one-halo term is large: the halo model recovers the IHL's cross-spectrum completely and the IGL's up to a residual that the
nonlinear two-halo term supplies. The IHL's auto-spectrum lies below every prediction, most at the lowest multipoles, the residual noted at
the end of Sec.~\ref{sec:bands} rather than an effect of nonlinearity.

The shading of Fig.~\ref{fig:nonlincross} shows where the detectability of
each cross-spectrum comes from in one simulated deep field, under the statistic of Sec.~\ref{sec:shen} with each component
alone in the covariance and the noise of the SPHEREx deep survey and of the Simons Observatory (SO)~\cite{Ade2019}
lensing map. For the lines half of it lies below $\ell \approx 400$--$700$, where
linear theory holds, so a linear model of one line's cross-spectrum with lensing is adequate. For the IGL and the IHL half of it
comes from $\ell > 700$, where the maps exceed linear theory, so their cross-spectra need at least the halo model.
A forecast of the lines is a third case, since they are measured
against the fluctuations of the IGL. Some forecasts, Ref.~\cite{Shen2025} among them, leave the shot noise out of
the auto-spectra; we keep it, as a population of point sources requires. In Paper~I~\cite{PaperI}, a no-shot linear
forecast doubles the detection of the summed lines over the SO footprint in the Limber approximation, from $7.3\sigma$ to
$15.0\sigma$, while a linear forecast that keeps the shot noise gives $6.2\sigma$. In the maps the IGL auto-spectrum follows linear theory only
below $\ell \approx 200$. It stands about $15\%$ above it over $\ell \approx 250$--$800$ (Sec.~\ref{sec:autos}) and is mostly shot
noise above $\ell \approx 1000$, so a no-shot linear forecast misses both.

These numbers carry over to the sky only in part, since the maps have no satellite galaxies and no diffuse IHL (Sec.~\ref{sec:samecons}), and the linear bias is the measured average over
$k < 0.3\hMpc$, which already absorbs part of the scale dependence.

\section{Filtering and detectability}
\label{sec:filtered}

We now apply the high-pass filter of Paper~I~\cite{PaperI} to the maps, compare the filtered spectra with the
prediction, and then compute the detectability statistic of Ref.~\cite{Shen2025}. The filter subtracts from each
LoS the part of its spectrum that varies slowly with wavelength. Section~\ref{sec:sim_th} tested the signal's level
in each channel; the filter removes that level and leaves the change from one channel to the next, which in these
maps comes from the lightcone: how each component evolves through the shells, and where the band ends.

\subsection{The filtered cross-spectra}
\label{sec:filtcross}

A SPHEREx channel is a wavelength and not a distance, so the modes removed along a LoS need a radial coordinate, and
we use the one of Paper~I~\cite{PaperI}: the comoving distance $\chi_i$ at which H$\alpha$ falls in channel $i$. Over
our 78 channels these distances span $6360$\,Mpc, so that the fundamental wavenumber of the band is $k_F \approx
9.9\times10^{-4}\Mpcinv$. The high-pass filter at the scale $\Lambda$ is then the projection
\begin{equation}
  \mathsf{H} = \mathbb{I} - QQ^{\sf T},\qquad m = \lfloor\Lambda/k_F\rfloor,
  \label{eq:filter}
\end{equation}
\begin{equation}
    Q = {\rm orth}\bigl\{1,\ \cos(jk_F\Delta\chi_i),\ \sin(jk_F\Delta\chi_i)\bigr\}_{j=1}^{m},
\end{equation}
with $\Delta\chi_i = \chi_i - \chi_{\min}$, which subtracts from the 78 channel values of every LoS their
least-squares fit by the constant and the $m$ lowest cosine and sine modes of the band, $2m+1$ modes in all.
Evaluated at the channels, these modes are vectors over the channels. Since the channels are unevenly spaced in
$\chi$, they are not orthogonal, and ${\rm orth}$ denotes their orthonormalization; Paper~I~\cite{PaperI} gives the
construction in full. Any quantity carrying a channel index is filtered by $\mathsf{H}$ on that index, applied, as in Paper~I, to the intensity
$I_\nu$, so that the filtered
cross-spectrum is $\sum_{i'}\mathsf{H}_{ii'}C_\ell^{\kappa I_{i'}}$. Filtering is linear and $\kappa$ is one map for
every channel, so we apply it to the measurement and to the prediction alike, and summarize each filtered
cross-spectrum at a given multipole by its rms over the channels.

The projection removes $2m+1$ linear combinations of the 78 channels outright, 21 of them at
$\Lambda = 0.01\Mpcinv$ and all 78 by $\Lambda \approx 0.038\Mpcinv$, beyond which no radial mode is left to filter. This limit is
set by the spacing of the SPHEREx channels, not by our maps. For the 102 channels of SPHEREx it lies at
$\Lambda \approx 0.040\Mpcinv$, close to ours, since in both cases it equals $\pi$ divided by the mean distance
between neighboring channels, about $80\,{\rm Mpc}$. The number of combinations that must be removed is set by the
IGL, which occupies only a few of them.
Its channel covariance leaves $1$--$3\times10^{-4}$ of its power outside its six leading eigenmodes, which are smooth functions
of the channel, for the total IGL of our maps as for the unresolved IGL of Paper~I~\cite{PaperI}, which we evaluate on the same channels. The nine combinations removed at $\Lambda =
0.0042\Mpcinv$, and the 21 at $0.01\Mpcinv$, therefore exceed what the IGL occupies, so the scales used here
cover what an analysis of the real survey would need. The same $\Lambda$ removes a different
fraction of the radial modes in Ref.~\cite{Shen2025}, whose $k_F$ follows the radial span of their slices: their
SPHEREx Ly$\alpha$ slices span $1080\,{\rm Mpc}$ in fifteen bins, so that a projection on their bins
would remove three of their fifteen combinations at $\Lambda = 0.01\Mpcinv$, against our 21 of 78, and would empty them near
$\Lambda \approx 0.04\Mpcinv$. They instead filter the continuous profile before binning it, which costs one combination
of their bins at every $\Lambda$ (their App.~D7) and lets them reach $\Lambda = 0.1\Mpcinv$.

An rms is positive, so noise adds to it in quadrature; we subtract the noise and quote the result, or a $2\sigma$
upper limit where the measurement is consistent with the noise. The maps carry no instrument noise, so this noise is the sample variance
of one $100\,{\rm deg}^2$ patch, which we take from the Gaussian error of Eq.~\eqref{eq:knox}
carried through the same filter, or, on the mean of the twenty realizations, from their scatter.

The filter treats that noise differently from one component to the next. Every channel of a smooth component sees
the same chance correlation of $\kappa$ with the LSS, which changes little from one channel to the next, so
the filter removes the noise together with the smooth part of the signal. A line's noise is its own shot noise,
uncorrelated between channels, and the filter passes it untouched while removing most of the signal.

\begin{figure*}[t]
  \centering
  \includegraphics{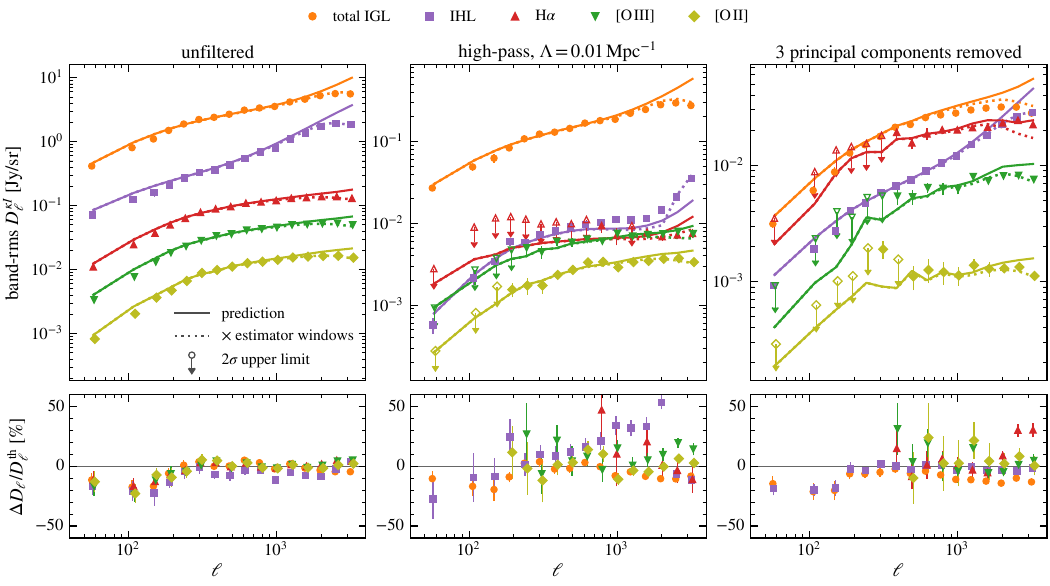}
  \caption{Cross-spectra of every component with the convergence, averaged over the twenty realizations and
  summarized at each multipole by their rms over the 78 channels (the band-rms of the axis label): without the filter (left), after it at
  $\Lambda = 0.01\Mpcinv$ (middle), and after the removal of three principal components of the channel covariance of
  the maps (right; Sec.~\ref{sec:filtcross}). Upper panels: the measured rms with its noise subtracted (points, with
  jackknife errors; the noise is the scatter between realizations), our prediction (solid) and the same multiplied
  channel by channel by the estimator's windows (dotted). Open symbols with arrows are $2\sigma$ upper limits,
  where the measurement is consistent with the noise. \emph{Lower panels}: the fractional difference, in percent, of the same
  measurements from the windowed prediction, drawn where the measurement is above the noise. After the filter the IHL
  and the three lines fall within a factor of a few of one another, so we read the comparison in the lower panels; the components are
  offset slightly in $\ell$ so that their error bars do not overlap. Averaging twenty $100\,{\rm deg}^2$ patches
  leaves the noise of about $2000\,{\rm deg}^2$.}
  \label{fig:filtkxens}
\end{figure*}

One patch therefore measures the filtered continuum and almost nothing of the lines. At $\Lambda = 0.01\Mpcinv$ the IGL keeps
about $6\%$ of its cross-spectrum, in the measurement and in the prediction alike (Fig.~\ref{fig:filtkxens}, left and
middle columns, for the mean of the twenty realizations), and is detected at
$5$--$19\sigma$ in eleven of the sixteen bins. The IHL keeps $0.5$--$2\%$ and is detected only in the two highest bins. The filtered H$\alpha$ and \OIII{} sit at the noise floor, with the prediction below it, while \OII{}, whose prediction exceeds the floor above $\ell \approx 1000$, is detected in the two highest bins. The fraction of the total IGL agrees with
that of Paper~I's unresolved IGL on the 102 SPHEREx channels, $5.9$--$6.2\%$ over the same multipoles, and the IHL
survives worst of all there too, since its cross-spectrum is the smoothest function of \mbox{distance} across the band. Every component but the IGL survives better here than there, the IHL by a factor of $2$ to $6$ and the lines by $1.3$ to $4.7$,
because our channel axis is the part of the SPHEREx band in which H$\alpha$ comes from the snapshots,
$1.01$--$4.56\um$ instead of $0.75$--$5\um$, and a profile cut where it is still high leaves a step that the filter
keeps (Sec.~\ref{sec:threemodels}). H$\alpha$ therefore survives here as well as the IGL or
better, while on the SPHEREx band it survives three times worse.

Averaging the twenty realizations, which is equivalent to reducing the noise to that of a $2000\,{\rm deg}^2$ survey,
lifts the lines above that floor, since their shot noise averages down as $1/\sqrt{20}$. The realizations are isometries of one
box and share its large-scale modes, so the noise they have in common does not average down, but that is the part the filter
removes in any case.

On the mean the filtered lines are measured. At $\Lambda = 0.0042\Mpcinv$, a weaker filter than that of
Fig.~\ref{fig:filtkxens} and not drawn there, H$\alpha$, \OIII{} and \OII{} are detected in nine, twelve
and thirteen of the sixteen bins, since only nine of the 78 combinations are removed there, against 21 at $\Lambda =
0.01\Mpcinv$. They agree with the prediction within about $2\sigma$, apart from \OIII{} at $\ell \approx 2500$, which
stands $16\%$ above it, a $4\sigma$ excess. The agreement needs a prediction built channel by channel from the
construction. It holds the growth of every component and the spectrum of the IGL at each snapshot, treats the shells
as independent realizations, builds each channel that straddles a shell boundary from both shells, and confines the
emitters to the H$\alpha$ range of the maps (Sec.~\ref{sec:samecons}). These details matter because the filter keeps
exactly what a comparison of single channels cannot see.

The two continuum components do not agree as well. At the filtering of Fig.~\ref{fig:filtkxens}, $\Lambda =
0.01\Mpcinv$, the IGL agrees with the prediction over $\ell \approx 240$--$780$ and lies $5$--$11\%$ below it above
$\ell \approx 1000$, while the IHL rises from
agreement at $\ell \approx 240$ to $1.5$ times the prediction at $\ell \approx 2000$ before falling below it in the
two highest bins. Unfiltered, channel by channel, the IGL sits within a few percent of the prediction and the IHL $5$--$8\%$ below.
The filter leaves a few percent of a smooth component's amplitude. It removes the overall offset and keeps the
part of the mismatch that varies quickly along the band, namely a drift of a few percent from the
blue end of the band to the red, which leaves the unfiltered rms untouched and sets the filtered one. The surviving
residual therefore need not share the sign of the unfiltered offset, and for the IHL it does not. The IHL is also
the component of which the filter leaves least, $0.5$--$2\%$ of its amplitude against the IGL's $6\%$, so the same
percent error in its predicted shape appears several times larger. We have not explained why its excess grows with multipole.
After the filter, the comparison tests the predicted shape along the band rather than its amplitude. The lines pass
within their errors, which the filter widens to between $5$ and $30\%$, apart from the \OIII{} excess at $\ell \approx 2500$, while at
$\ell \approx 1000$ the IGL lies six standard deviations below the prediction and the IHL four above.

Unfiltered, every component agrees with the prediction, apart from the IHL at $\ell \approx 1000$--$2000$
(Sec.~\ref{sec:bands}), and all five fall below it together at the lowest multipoles (Fig.~\ref{fig:filtkxens}, lower left), which an error in one component's model would not do. This common shortfall is consistent with the
leakage of Sec.~\ref{sec:estimator}, which the twenty realizations cannot settle
(Sec.~\ref{sec:bands}). We therefore read the comparison above the four lowest bins.

Paper~I~\cite{PaperI} finds that removing the leading principal components of the channel covariance keeps more of
the line signal than the high-pass filter. We test this on the maps and estimate the components as an analysis would,
from the measured channel covariance of the total map of one realization, with the white noise of the SPHEREx deep
fields added (Sec.~\ref{sec:shen}) and the multipoles summed with their mode counts. The right column of
Fig.~\ref{fig:filtkxens} shows the mean of the twenty realizations after three components are removed. The removal
keeps $0.9\%$ of the IGL, against $6\%$ for the high-pass filter at $\Lambda = 0.01\Mpcinv$. H$\alpha$ keeps
$16$--$21\%$ of its amplitude, two to three times as much as under the high-pass filter, and \OIII{} and \OII{} keep
$8$--$14\%$. What remains of the IGL and the IHL is still $1.2$ to $1.8$ times the summed lines, so the IGL, the IHL and the lines must still be separated by their templates. The IHL agrees with the prediction within $4\%$, and the IGL lies
$10$--$14\%$ below it above $\ell \approx 1000$, as after the high-pass filter. The lines are detected above $\ell
\approx 400$ and agree with the prediction within about $2\sigma$, apart from H$\alpha$ in the two highest bins, which
stands $30\%$ above it.

In the detectability statistic of Sec.~\ref{sec:shen}, with the amplitudes of the IGL and the IHL fitted beside that
of the summed lines, the removal of three principal components keeps $76\%$ of the signal-to-noise ratio of the summed lines over the two
deep fields, and the high-pass filter at $\Lambda = 0.01\Mpcinv$ keeps $27\%$. The components also leave a tenth of
the continuum power that the high-pass filter leaves. Estimated from one realization, they leave the fitted amplitude
of the lines in the other nineteen unchanged, $0.94 \pm 0.03$ against $0.95 \pm 0.02$ without the removal. The
continuum of our maps is as coherent across the channels as that of Paper~I~\cite{PaperI}, however, so this tests
how well the components are estimated, not the spectral diversity of real galaxies. On a semi-analytic lightcone
whose galaxies each carry their own star-formation history, metallicity and dust, removing about twenty components of
91 channels gave Ref.~\cite{Roy2026a} the best balance between keeping the lines and suppressing the continuum at every
multipole they tested, $\ell = 2000$--$14\,000$. In our maps the spectrum of a halo is set by its mass and redshift,
and the scatter rescales all wavelengths alike, so three components leave $5\times10^{-4}$ of the continuum power. How
many components the cross-spectrum with lensing needs on maps like theirs has not been tested.

\subsection{The detectability statistic on the maps}
\label{sec:shen}
\begin{figure*}[t]
  \centering
  \includegraphics{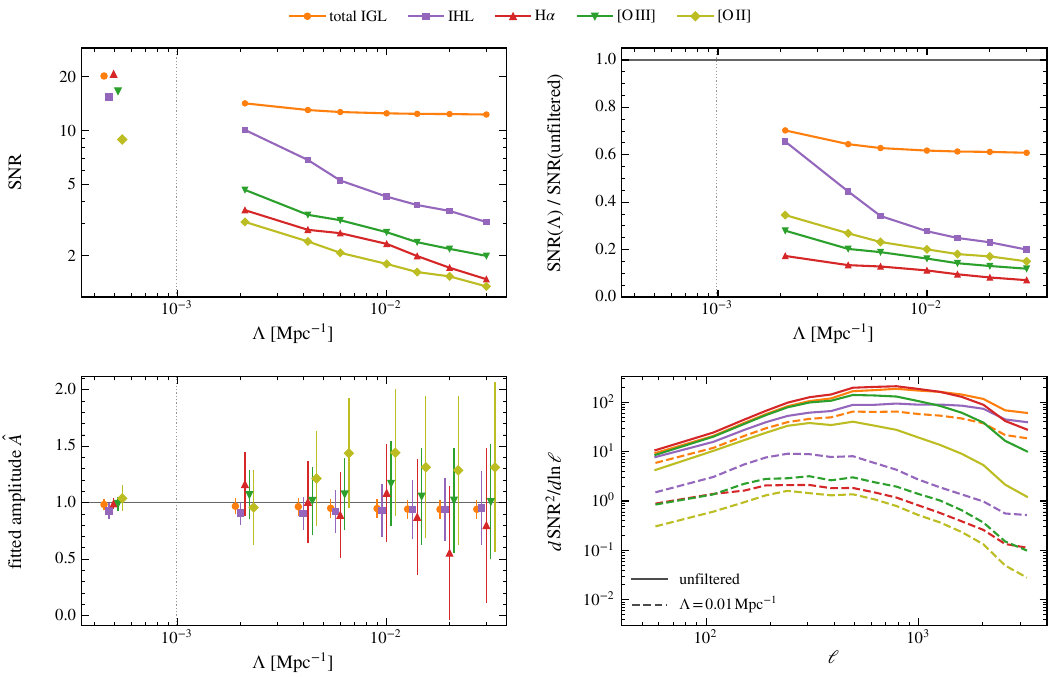}
  \caption{The detectability statistic of Ref.~\cite{Shen2025} on one $100\,{\rm deg}^2$ patch with the 78 channels as radial slices.
  The curves are the prediction; only the amplitudes (bottom left) are measured. Every panel carries the deep-field white
  noise and the SO lensing noise. Top left: the signal-to-noise ratio against the high-pass scale $\Lambda$; the symbols at
  $k_F/2$ are the unfiltered values. Top right: the fraction of the
  signal-to-noise ratio that survives the filter. Bottom left: the amplitude $\hat A$ of the windowed prediction
  fitted to the maps, with its error $\sigma_{\hat A}$; the components are offset slightly in
  $\Lambda$, and the leftmost points are unfiltered. In these three panels the vertical dotted line marks $k_F$, the
  smallest $\Lambda$ at which the filter removes more than the channel mean. Bottom right: where the squared
  signal-to-noise ratio accumulates in $\ell$, unfiltered (solid) and at $\Lambda = 0.01\Mpcinv$ (dashed).}
  \label{fig:shensnr}
\end{figure*}

We now turn to the statistic with which Ref.~\cite{Shen2025}
quantifies detectability. There, the cross-spectrum is measured in every radial slice, the slices are
combined at each multipole with the inverse of their full covariance, and the result is summed over multipoles. Paper~I~\cite{PaperI} applies this statistic to its model of every SPHEREx component, and here we apply it to the
maps, with the 78 channels as the slices. The covariance $\Sigma_\ell/2N_\ell$ of the measured cross-spectrum vector, the statistic
${\rm SNR}^2 = \sum_\ell 2N_\ell\, O_{\ell,f}^{\sf T}\,\Sigma_{\ell,f}^{+}\,O_{\ell,f}$ and the filter are those of Paper~I~\cite{PaperI}, with $O_{\ell,i} = C_\ell^{\kappa I_i}$ the cross-spectrum of channel $i$ and $N_\ell$ the mode
count of Eq.~\eqref{eq:knox}, in which a mode and its conjugate count once. The subscript $f$ marks the quantities
filtered by Eq.~\eqref{eq:filter} on their channel indices, with $\mathsf{H} = \mathbb{I}$ unfiltered, and $^{+}$ is
the pseudo-inverse on the non-null subspace.
To the statistic we add the amplitude of the prediction fitted to the measured cross-spectrum $\hat O$,
\begin{equation}
  \hat A = \frac{\sum_\ell 2N_\ell\, O_{\ell,f}^{w\,\sf T}\,\Sigma_{\ell,f}^{+}\,\hat O_{\ell,f}}
                 {\sum_\ell 2N_\ell\, O_{\ell,f}^{w\,\sf T}\,\Sigma_{\ell,f}^{+}\,O^{w}_{\ell,f}},
  \label{eq:simsnr}
\end{equation}
where $O^w$ is the predicted cross-spectrum times the estimator's windows (Sec.~\ref{sec:crosses}), the template fitted to the maps.
The error $\sigma_{\hat A}$ is the inverse square root of the denominator, that is, the inverse signal-to-noise ratio of
the windowed template, which the windows leave within $5\%$ of ${\rm SNR}^{-1}$.

The statistic does not change under any invertible transformation of the data. A smooth high-pass filter
suppresses the low-$k_\parallel$ modes without removing them, so it would leave the statistic unchanged. For this
reason, the filter has to be a projection.

A covariance estimated from one patch has rank at most $2N_\ell$, below 78 for $\ell \lesssim 250$ and too low to
invert reliably well above it, so every element of $\Sigma_\ell$ is taken from theory. The cross-spectra and the channel covariances come from the prediction of Sec.~\ref{sec:samecons}, with the box's own clustering in the two-halo terms and a \mbox{diagonal} \mbox{covariance} for each line, whose channels are disjoint shells, which neglects the weak correlation between neighbors. For $C_\ell^{\kappa\kappa}$ we take the halofit~\cite{Smith2003,Takahashi2012} Limber spectrum of the lightcone, and for $N_\ell^\kappa$ the baseline minimum-variance reconstruction noise of SO~\cite{Ade2019}, drawn against the signal in Fig.~1 of Paper~I~\cite{PaperI}. The white noise is set by the SPHEREx deep-field surface-brightness sensitivity measured in flight, read per channel from Fig.~19 of Ref.~\cite{Bock2025}.

Figure~\ref{fig:shensnr} gives the result. Unfiltered, the lines are detected at $9$--$21\sigma$ in one
patch, the IGL at $20\sigma$ and the IHL at $15\sigma$, and the simulated lines are bright enough to be limited by sample variance in their best channels: at
$2.2\um$ and $\ell \approx 1000$ the white noise is $0.07$ of H$\alpha$'s own power, against $0.003$ of the IGL's.
More depth would not help there, since the signal-to-noise ratio is set by the number of modes and by
the correlation with $\kappa$. At $\Lambda = 0.01\Mpcinv$ the high-pass filter leaves the IGL $62\%$ of its unfiltered
significance, the IHL $28\%$ and the lines $11$--$20\%$, so that after it the total IGL leads, as the unresolved IGL does in Paper~I's
forecast. These fractions are far larger than the fractions of the amplitude that survive, $6\%$ for the IGL
(Sec.~\ref{sec:filtcross}), because the filter removes the error with the signal. The IGL's cross-spectrum is limited by
its own sample variance, whose channel covariance is as smooth along the band as the signal, so the projection
takes out both and the error falls by a factor of about ten. Repeating the statistic with no noise of any kind separates two causes of these losses. The filter then
leaves the IGL $71\%$ and H$\alpha$ $11\%$, close to their values with noise, so the gap between those two follows
the shape of their cross-spectra along the channel axis and not the noise floor, the white noise of the channel
maps. For the IHL, \OIII{} and \OII{} the same
filter leaves $84$, $45$ and $56\%$ without noise, so the white noise costs them the rest. It is white along the
channel axis, whereas their signal is not, and the projection removes 21 of the 78 combinations. We use the noiseless
statistic only for this comparison. It is idealized, and for the IHL threshold-dependent: without noise only 30 of
its 78 channel combinations carry variance above the threshold of the pseudo-inverse, $10^{-10}$ of the largest
eigenvalue, at $\ell \approx 1000$.

The amplitudes fitted to the maps are consistent with the prediction: unfiltered and after filtering, every one
lies within $1.1\sigma$ of unity. The statistic combines the channel-by-channel cross-spectra of
Sec.~\ref{sec:filtcross}, weighted by the inverse of their covariance. The offsets of the IGL and the IHL that the mean
of the twenty realizations resolves there amount to about one standard deviation of one patch, whose errors are
$\sqrt{20}$ times larger.

\begin{figure*}[t]
  \centering
  \includegraphics{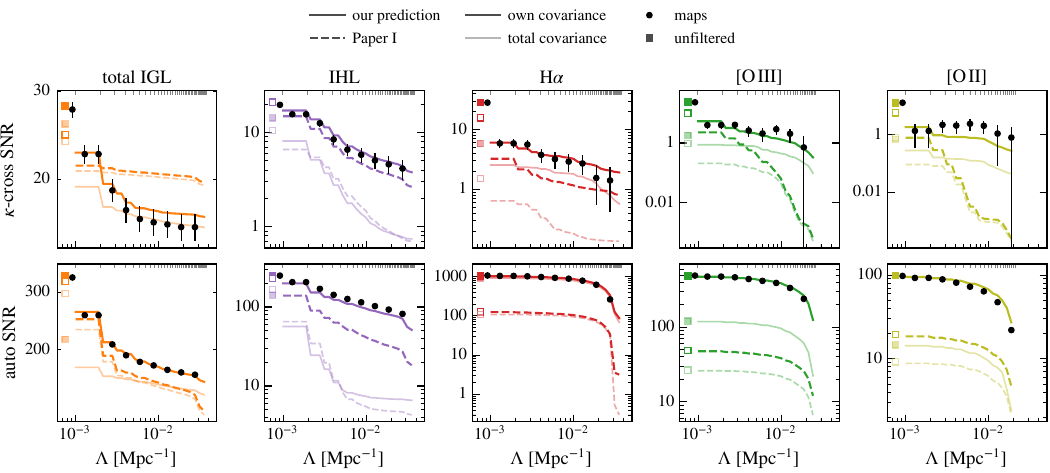}
  \caption{Detectability against the high-pass scale $\Lambda$ for the $\kappa$-cross-spectrum (top row) and the auto-spectrum
  (bottom row) of each component (columns), on the maps' 78 channels ($1.01$--$4.56\um$; for \OIII{} and \OII{}, the 53 and 42 channels their emitters reach) and sixteen multipole bins, over
  $200\,{\rm deg}^2$ with the deep-field white noise and the SO lensing noise. Solid: our prediction;
  dashed: Paper~I's model~\cite{PaperI}, with its unresolved IGL, restricted to the lightcone. Dark: each component's own
  covariance; faint: the covariance of the total intensity, the five components summed. Black points: the maps, as the
  fitted amplitude of our template times that template's signal-to-noise ratio (own covariance), with
  $1\sigma$ errors for $200\,{\rm deg}^2$. Squares at the left edge: the unfiltered values of the curves (filled: our prediction; open: Paper~I's model; faint: total covariance), drawn at a nominal $\Lambda =
  7.5\times10^{-4}\Mpcinv$ beside the maps' unfiltered point. The gray ticks on the upper edge mark the multiples of
  $k_F$, where every curve steps.}
  \label{fig:snrthree}
\end{figure*}

\begin{figure*}[t]
  \centering
  \includegraphics{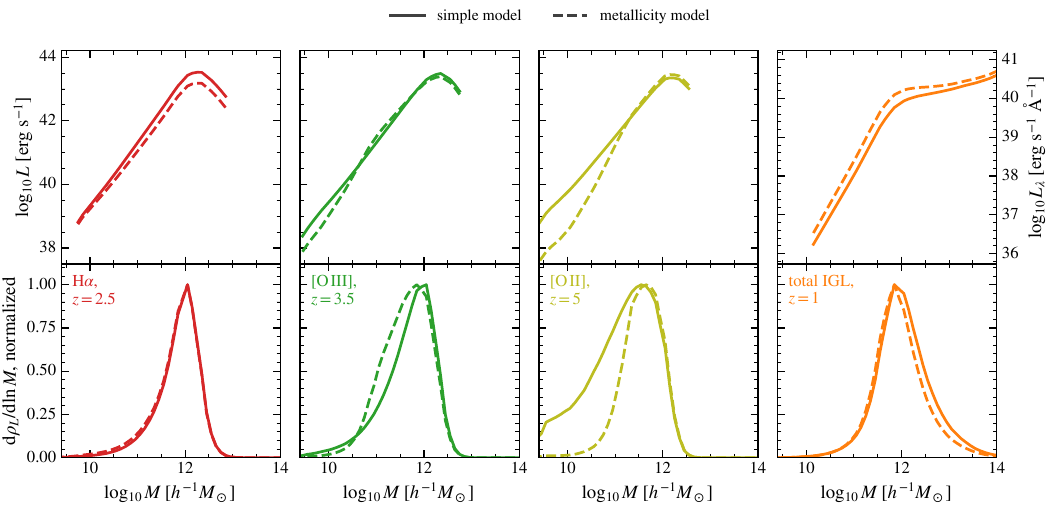}
  \caption{The luminosity per halo in the two line models, for H$\alpha$ at $z = 2.5$, \OIII{} at $z = 3.5$, \OII{} at
  $z = 5$ and the IGL at $z = 1$, at rest-frame $1.10\um$ (columns), the snapshots whose shells the $2.2\um$ channel sees. Top row: the
  luminosity of a halo along the median relation, in the simple model (solid) and the metallicity model (dashed); the
  IGL per \AA, on the right-hand axis. Bottom row: the luminosity density per $\ln M$, normalized to its peak.}
  \label{fig:linemodelmass}
\end{figure*}

\begin{figure*}[t]
  \centering
  \includegraphics{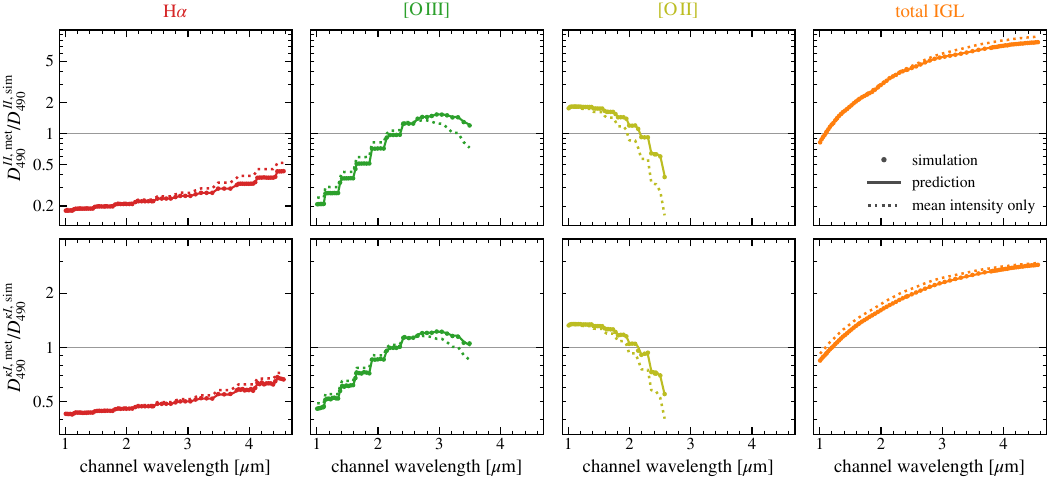}
  \caption{The metallicity model over the simple model, channel by channel at $\ell \approx 490$: the auto-spectra
  (top) and the $\kappa$-cross-spectra (bottom), for H$\alpha$, \OIII{}, \OII{} and the IGL (columns), with ``met'' and
  ``sim'' labeling the two models. Points: the ratio measured on the twenty paired realizations, with jackknife
  errors. Solid: the ratio of the two predictions. Dotted: the ratio of the two mean intensities, squared for the
  auto-spectrum, which is what a change of the mean intensity alone would give.}
  \label{fig:linemodels}
\end{figure*}

\subsection{Three models side by side}
\label{sec:threemodels}

Lastly, we compare the maps, our prediction and Paper~I's model~\cite{PaperI} on the same channels, depth, sky area
and statistic (Fig.~\ref{fig:snrthree}), over $200\,{\rm deg}^2$, the area of the two SPHEREx deep fields, with
Paper~I's model restricted to the lightcone's $z = 0.25$--$6.25$ and evaluated in Limber, without redshift-space distortions. For the auto-spectra the statistic is the Fisher
signal-to-noise ratio of each component's channel covariance, as in Paper~I.

Unfiltered, the two descriptions of the IGL agree to within $20\%$, while those of the lines differ by factors of two
to ten. Paper~I's unresolved IGL and the maps' total IGL share only the halo model, yet give nearly the same cross-spectrum
detectability, and the two IHL models, which share their luminosity calibration, agree to within $1.5\%$. The maps' lines
give about twice Paper~I's cross-spectrum detectability and five to ten times its auto-spectrum detectability, because they are brighter: by
factors of three to five at $2.2\um$, and by up to an order of magnitude across the band. Four differences between Paper~I's line model and the simple model, which we list in Sec.~\ref{sec:linemodels}, contribute to that brightness. A per-line detectability is therefore uncertain by a factor of two in the cross-spectrum and up to ten in the auto-spectrum.

After the filter, the two descriptions of the IGL and the two IHL models no longer agree, for different reasons.
At $\Lambda = 0.01\Mpcinv$ the IHL keeps $21\%$ of its unfiltered detectability in Paper~I's model and $28\%$ in
ours. The difference lies in the channel covariance and not in the cross-spectrum, since exchanging the cross-spectra of the two models moves these fractions by $0.003$, while exchanging their covariances moves them from $0.21$ to $0.28$. The maps' IHL
covariance exceeds Paper~I's by a factor growing from $1.5$ at $\ell \approx 60$ to $6$ at $\ell \approx 3200$,
because the maps paint the IHL as a point carrying the $0.3$\,dex luminosity scatter, whereas Paper~I keeps the diffuse
profile (Appendix~\ref{app:ingredients}), so its filtered signal stands further above the white-noise floor. For the
IGL, the two statistics differ: the auto-spectrum keeps $49\%$ in Paper~I's model and $52\%$ in ours, and the cross-spectrum $84\%$ and $62\%$. The auto-spectrum
statistic is built from the channel covariance alone, and the two covariances have nearly the same structure across
channels, leaving $1.1\times10^{-3}$ and $4.4\times10^{-4}$ of their power outside the three leading eigenmodes of their channel correlation matrices. The cross-spectrum
statistic weights a profile against that covariance, and the two profiles run opposite ways across the band: at $\ell
\approx 1000$ Paper~I's cross-spectrum is twice as large at the red end of our channel range as at the blue, and the maps'
twice as large at the blue end as at the red, since from $1$ to $3\um$, in $I_\nu$, Paper~I's unresolved IGL rises
and the maps' total IGL falls. We have not reduced the gap between $84$ and $62\%$ to a single feature of those profiles.

The black points of Fig.~\ref{fig:snrthree} test whether the maps reproduce the prediction. Unfiltered, the fitted amplitudes lie
within $7\%$ of unity for the cross-spectra and within $2\%$ for the auto-spectra, whose template carries the pixel window on the
clustering term of every channel (Sec.~\ref{sec:autos}). After the filter the cross-spectrum amplitudes scatter more widely but stay within
$1.8\sigma$ of unity; the filtered line templates have a signal-to-noise ratio between a fraction of unity and a few.
These errors are those of $200\,{\rm deg}^2$. Not all auto-spectrum amplitudes stay as close. At $\Lambda = 0.01\Mpcinv$, where their
errors are $1.5\%$ or less, the IHL's is $1.16$ and those of H$\alpha$, \OIII{} and \OII{} are $0.99$, $0.97$ and
$0.88$, while the IGL's stays between $0.98$ and $1.01$ at every $\Lambda$. We have not traced these departures. The IGL's fitted auto-spectrum amplitude is the most demanding of these tests,
and it is a test of the shape of the template across the channels rather than of its level, since the IGL's channel covariance is nearly rank one, and half of the weight of the fit falls on the second and third eigenmodes, which carry well under a
percent of its power. It reaches unity only because the template reads each shell's spectrum at the snapshot, as the
maps do; a template whose spectrum evolves within the shell puts power into those same eigenmodes and fits
$13\%$ high.

The faint curves of Fig.~\ref{fig:snrthree} put every component's variance into the covariance. In the cross-spectrum this costs the lines a factor
of five to eighteen unfiltered, through the IGL's variance in the line channels. After the filter the cost
falls to $1.8$ for the maps' H$\alpha$ and $6.4$ for Paper~I's, because the IGL's variance is smooth
across channels: the filter removes that variance along with most of the line's cross-spectrum signal.

The filtered results carry two features of this construction that a forecast does not share. The maps are built
from snapshots, so that within each shell the halos, their clustering and their spectra are those of one epoch, and
the prediction holds them there too; a theory that lets them evolve through a shell carries a sawtooth along the
channel axis, which no statistic averaged over the channels can see and a high-pass filter keeps in full
(Appendix~\ref{app:ingredients}). The channel axis also begins at $1.01\um$, where the H$\alpha$ cross-spectrum at
$\ell \approx 300$ is still $45\%$ of its peak across the band, and the projection keeps that step: on this channel range
Paper~I's filtered H$\alpha$ detectability at $\Lambda = 0.01\Mpcinv$ is $1.17$, against $0.21$ on the full SPHEREx
band, whose blue edge at $0.75\um$ sits where the cross-spectrum is $5\%$ of the peak. The filtered line cross-spectra of
this section are therefore a comparison of models and a consistency test, not a forecast. Beyond these, no curve carries
zodiacal or Galactic light, the total covariance omits the covariances between components, Paper~I's IGL is masked whereas the maps' is not, and neither deep field lies inside the $16\,000\,{\rm deg}^2$ footprint adopted for SO
lensing forecasts~\cite{Ade2019}, so the $200\,{\rm deg}^2$ curves assume SO lensing noise over both, as
Ref.~\cite{Shen2025} does. The forecast for SPHEREx, with its
survey assumptions, is that of Paper~I~\cite{PaperI}.

\section{A second line model}
\label{sec:linemodels}

In this section, we repaint the twenty realizations with the metallicity model and measure how the auto- and
cross-spectra change across the band. Each halo
takes a stellar mass from the median relation of the UniverseMachine~\cite{Behroozi2019}; that stellar mass sets its
gas and stellar metallicity through the mass--metallicity relation of Ref.~\cite{Curti2020}, evolved with redshift as
in Ref.~\cite{Sanders2021}; and the ionization parameter of the gas falls with that metallicity as in
Ref.~\cite{Dopita2006}. The three lines then no longer keep fixed ratios, and none of them is proportional to the
star-formation rate.

The metallicity model adds no scatter of its own, and both paintings carry the same $0.3$\,dex scatter at fixed
mass. The two paintings share the halos, the isometries, the
$\kappa$ maps and the scatter draws, so that their difference is the line model alone, measured realization by
realization with errors far below the sample variance of either map. The IHL does not depend on the line model and is
identical in the two paintings.

\subsection{The two luminosity--mass relations}
\label{sec:repaint}

In Fig.~\ref{fig:linemodelmass} we show the luminosity of a halo in
the two models and where the light comes from in halo mass. The metallicity model multiplies each halo's luminosity by a factor that depends on
its mass, and a map changes only as much as that factor varies across the halos that carry the light. For
H$\alpha$ that factor falls by about a half from $10^{10}$ to $10^{13}\Msunh$, but is nearly constant over the
decade of mass that carries its light, so the H$\alpha$ map is close to a dimmed copy and its bias barely moves. The
contrast with the oxygen lines is the one to expect. H$\alpha$ is a recombination line and counts the ionizing
photons of the stars, on which the metallicity acts only through the hardness of the stellar spectrum, while the
oxygen lines are collisionally excited and so depend on the abundance of oxygen and on the ionization parameter
directly, both of which follow the halo's mass here.
The metallicity model almost switches off the low-mass \OII{} emitters at $z = 5$ while brightening the massive
halos, moving the \OII{} light up in mass by a factor of two, so that the \OII{} map is repainted rather than rescaled. \OIII{}, on
the other hand, brightens near $10^{11}\Msunh$ and dims in more massive halos, and its light moves slightly down. The IGL per halo rises toward
low mass by about a factor of two over the masses that carry its light, which lowers its bias by a few percent;
the $2.2\um$ IGL map sums the shells from $z = 0.25$ to $6.25$, which dilutes the change and leaves it close to a
rescaled copy.

\subsection{The change across the band}
\label{sec:linemodelband}

In Fig.~\ref{fig:linemodels} we show the ratio of the auto-spectra and the $\kappa$-cross-spectra of the two models at $\ell \approx 490$. The mean intensity changes by large factors across the band. H$\alpha$
falls to $0.43$--$0.72$ of its simple-model intensity, the least at high redshift, because the stars are no longer at the lowest metallicity and emit fewer
ionizing photons. \OIII{} and \OII{} trade places with redshift: \OIII{} rises from half its simple-model intensity at $z \approx 1$
to slightly above it at $z \approx 4.5$, while \OII{} falls from a third above it at $z \approx 2$ to well below it by $z \approx 6$. Galaxies are
metal-poorer at high redshift, so their ionization parameter is higher, which favors \OIII{} over \OII. The IGL
changes most, from $0.92$ of its simple-model intensity in the bluest channel to $2.94$ in the reddest, since metal-rich stars are redder and the long-wavelength channels see the rest-frame near infrared.

Once the change of intensity is divided out, the $\kappa$-cross-spectrum of a line changes as its bias does: the measured
change of the cross-spectrum, divided by the change of the mean intensity, equals the ratio of the two models' biases
measured in the box (Fig.~\ref{fig:biaslum}, bottom) to within a few percent in every channel. The auto-spectra instead follow the shot noise on small scales, which
weights the brightest halos most, so that the \OII{} auto-spectrum, whose light moves into fewer and more massive halos,
changes by more than its intensity squared, and increasingly so toward high multipoles.

The ratio of the two predictions reproduces the measured ratio channel by channel, to within $2\%$ in every channel
for H$\alpha$, \OIII{} and the IGL and within $4\%$ for \OII{}, auto- and cross-spectrum alike, where each prediction is built from its own model's luminosities and box-measured biases. The
ratio of the two mean intensities alone does not (dotted in Fig.~\ref{fig:linemodels}): a change of the line model is
not a rescaling of the maps. The comparisons of Sec.~\ref{sec:sim_th}
therefore hold for the metallicity model as well, to within a few percent, bin for bin.

The detectability changes far less than the mean intensity. In the statistic of Sec.~\ref{sec:shen} each line's
unfiltered $\kappa$-cross-spectrum signal-to-noise ratio moves by at most $15\%$, although H$\alpha$'s signal falls by half:
where a line is best measured, its cross-spectrum is limited by its own sample variance, which falls with the signal.
After the high-pass they change by less than $30\%$. The IGL's unfiltered value does not move
and its filtered value rises by $42\%$, which its channel covariance would explain, though we have not tested it.

In the box, the correlation of each luminosity-weighted field with the matter, $r(k) = P_{wm}/\sqrt{P_{ww}P_{mm}}$
with the shot noise kept in $P_{ww}$ as a map has it, falls below $0.9$ between $k \approx 0.15$ and $0.4\hMpc$ in
the simple model, soonest for \OII{}. The metallicity
model moves \OII{} at $z = 5$ into the few massive, metal-richer halos, where its bias rises by a fifth and its shot
noise nearly triples. \OII{} becomes a more biased but noisier tracer, and its correlation with the matter falls by about a tenth on the
smallest scales measured, while \OIII{} and H$\alpha$ barely move.

We note two aspects of the line models. First, the \OII{} results at $z \approx 5$--$6$ rest on the
evolution of the mass--metallicity relation of Ref.~\cite{Sanders2021}, extrapolated beyond its $z \approx 3.3$, the
least certain input of the metallicity model. Second, the factor of up to three between the two models in a channel's
intensity measures how weakly that intensity is constrained, and neither model is a correction to the other: we quote both
rather than one with an error bar. Of the two, the metallicity model is the more physically motivated, since the
simple model holds every halo at one metallicity and, by the defaults of the code, at the lowest stellar metallicity
of the library with solar nebular gas (Sec.~\ref{sec:lim}). Measurements of the line emission at these redshifts will
settle which of them is closer to the sky (Sec.~\ref{sec:conclusions}).

The simulated lines are also three to five times brighter at $2.2\um$
than the lines of Paper~I~\cite{PaperI}, which are calibrated to the cosmic star-formation-rate density
(Sec.~\ref{sec:threemodels}). Four differences contribute. The first is the dust of the FSPS spectra, against the
differential extinctions of the line model of Ref.~\cite{Gong2017}. The others are the star-formation history of the
UniverseMachine~\cite{Behroozi2019}, against that of Ref.~\cite{HopkinsBeacom2006}; the maps' $0.3$\,dex
scatter; and the stellar metallicity of the simple model, which by itself raises H$\alpha$ by a factor of $2.4$ over a
solar-metallicity population. We have not combined these differences into a budget. The per-line detectabilities carry
this uncertainty until the line luminosities are anchored to observed luminosity functions. The agreement of the maps with the prediction, in contrast, does not depend on the line model.

\section{Conclusions}
\label{sec:conclusions}

We have built paired SPHEREx intensity maps and CMB-lensing convergence maps from the halos and the matter of one
$N$-body box, and compared every spectrum they contain with a prediction made from the same ingredients, with no
free parameter. With twenty realizations of a $10^\circ\times10^\circ$ patch to $z = 6.25$, we have tested the
machinery that forecasts of line-intensity maps crossed with lensing rely on: the halo model, the filter along the
LoS and the detectability statistic.

The prediction reproduces the maps, but only when it is built for them. The convergence agrees with its expected
spectrum to within $7\%$ at every multipole we measure. That expectation has to carry the discreteness of the
shells, the projection of each part of a shell at its own distance, and the transfer of power by the lattice, the
patch and the pixel. At $2.2\um$ the auto-spectra of the IGL and of the three lines then agree to $5\%$ and their
cross-spectra with $\kappa$ to $7\%$, and the agreement holds channel by channel across the band. The agreement also requires the clustering of the box itself. The halo field weighted by the IGL follows linear bias on the
nonlinear matter power, while the halos that carry the lines and the IHL are biased nonlinearly as well, the rarest the
most. The prediction does not reproduce the IHL: its cross-spectrum runs about a tenth low where it is best measured, which we
have not traced to any ingredient.

Linear theory describes the cross-spectrum of a single line with lensing to a few percent up to a multipole of about a
thousand. That covers most of the scales where the cross-spectrum is detectable. For the two continuum components it fails
already from a few hundred, through the one-halo term, and by a multipole of a thousand it is low by $30\%$ for the
IGL and by a factor of three for the IHL. A forecast without shot noise also misses the fluctuations of the
continuum against which the lines are measured.

The filter along the LoS and the detectability statistic of Paper~I both transfer to the maps. The
filtered line cross-spectra of the twenty realizations agree with the prediction to about $2\sigma$, apart from one \OIII{} bin at $4\sigma$, and the fitted
amplitudes of the statistic are consistent with unity before and after filtering. The filter keeps only what varies quickly from channel to
channel, so the filtered comparison tests the predicted shape rather than its level. The lines pass that test, while the
two continuum components depart from the predicted shape. On the same channels, the simulated lines are
several times brighter than those of Paper~I and give about twice its cross-spectrum detectability, while the maps' total IGL and Paper~I's unresolved IGL are equally detectable.

We have repainted the maps with a second line model, in which the metallicity follows the halo. It changes the intensity
of a channel by up to a factor of three, and the cross-spectrum detectability of the lines by only tens of percent. The detectability of a cross-spectrum limited by a component's own sample variance does not depend on the amplitude of that component, since the variance falls
with the signal. The prediction follows either model and reproduces its maps, so none of the tests above rests on
the choice between them.

Every disagreement that we traced was a defect of the construction rather than of the physics, and
Appendix~\ref{app:defects} describes two of them with the tests that catch them. Among the spectra, only a cross-correlation
reveals a misregistration between two maps, and no comparison of ours reveals an error in an input that the maps
and the prediction share.

This work can be extended in several directions. Maps of this kind can test what an analytic forecast cannot,
component separation among them. The ensemble is limited most by the large-scale modes that its
twenty realizations share, and a second, independent box would show how much of what we see at the lowest multipoles
is the cosmic variance of one box. A continuous lightcone would remove the steps at the shell boundaries, which a
high-pass filter keeps and which we have had to carry through every filtered comparison. Each halo now
carries its light at a single point. Satellite galaxies, light profiles within halos and a diffuse intrahalo light would
spread that light over the halo and change the one-halo terms of the spectra.

The level of the lines remains the largest uncertainty of all: our two models differ by up to a factor of two and a half
in a line's intensity. An observed luminosity function of H$\alpha$ at $z \approx 1$--$2$ would fix the bright end of the models, but a
line-intensity map also holds the light of the emitters below any flux limit. The cross-correlation of SPHEREx with a
spectroscopic sample of galaxies at the same redshifts measures instead the mean intensity times the bias of all
unmasked emitters, while its Poisson term measures the light of the cataloged galaxies that the mask leaves in the
map~\cite{ChengChang2021}. The first is
the combination that enters the cross-spectrum with lensing. Euclid's spectroscopic survey selects H$\alpha$ emitters
at $z = 0.84$--$1.88$ over its wide survey, and from $z = 0.41$ in its deep fields, which add a bluer
grism~\cite{Mellier2024}. Euclid's wide survey, about
$14\,000\,{\rm deg}^2$ of the sky that SPHEREx maps in full, has the area to measure the clustering term precisely,
whereas even the two SPHEREx deep fields together are limited by sample variance~\cite{ChengChang2021}. Its northern deep field, $20\,{\rm deg}^2$
at the north ecliptic pole~\cite{Mellier2024}, lies inside the SPHEREx north deep field~\cite{Bock2025,ChengChang2021}, where both
surveys reach about two magnitudes deeper than over the wide sky. There the cataloged galaxies carry more of the light,
and the Poisson term measures how much~\cite{ChengChang2021}.

\begin{acknowledgments}
We thank Tom Chiesura for his contribution at the early stages of the project. We also thank Abhishek Maniyar and Giulio Fabbian
for helpful discussions regarding the prospects of measuring the cross-correlations on
the observed maps. This work is supported by the Agence Nationale de la Recherche (ANR) under grant No.~ANR-23-CPJ1-0160-01.
The authors acknowledge the use of Claude Code (Anthropic) in this work, to implement some of the analysis codes and to
prepare the manuscript. They designed and directed the work, which builds on their earlier
pipeline, reviewed the code, verified all results, and take full responsibility for the paper's content.
\end{acknowledgments}

\appendix
\section{The halo catalog}
\label{app:halos}

\begin{figure}[tb]
  \centering
  \includegraphics[width=0.9\linewidth]{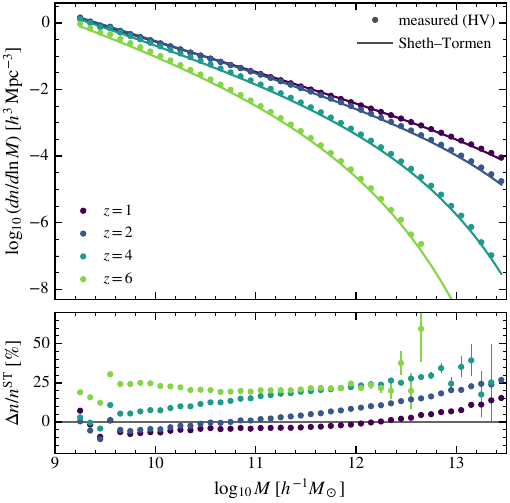}
  \caption{The halo mass function of the box at four snapshots, $z = 1$ to $6$ (points, colored from dark purple to
  light green), in bins of $0.1$\,dex, against the ST prediction (solid lines of the same color).
  \emph{Bottom}: the fractional difference, in percent, of the measurement from ST, with Poisson error
  bars; bins holding fewer than $20$ halos are left out. The factor $M$ cancels in that ratio, which is therefore
  the same for $n$ and for $M n$.}
  \label{fig:massfunction}
\end{figure}

In this appendix, we compare the abundance and the bias of the halos in the box with the ST
predictions~\cite{ShethTormen1999}. In Fig.~\ref{fig:massfunction} we show the measured mass function per \mbox{logarithmic} mass interval, $M\,n(M,z) = dn/d\ln M$, at four snapshots against ST: the two agree to within about $10\%$ below $10^{12}\Msunh$ at $z \le 2$, and the ST
prediction falls increasingly short toward high mass and high redshift, by tens of percent at $z = 4$--$6$;
Ref.~\cite{MoradinezhadDizgah2022}, whose maps are built on the same catalogs, reports the same trend. These offsets
enter the mean intensity and the shot noise of a map in full, so our prediction uses the measured abundance
(Sec.~\ref{sec:samecons}), whereas the luminosity-weighted bias is a ratio of two integrals over the abundance, in
which it largely cancels (Sec.~\ref{sec:bias}).

\begin{figure}[tb]
  \centering
  \includegraphics[width=0.9\linewidth]{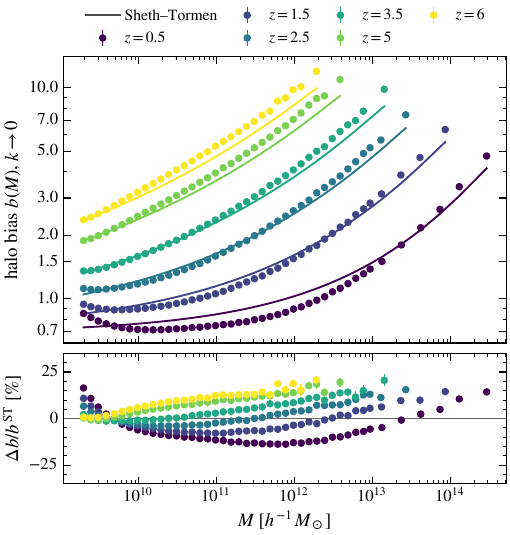}
  \caption{The large-scale halo bias in mass bins of $0.1$\,dex, wider at the massive end so that each holds at least
  2000 halos, at six snapshots, $z = 0.5$ to $6$ (points with their errors, colored from dark purple to yellow), against ST averaged over
  each bin (solid lines of the same color). \emph{Bottom}: the fractional difference, in percent, of the measurement from ST.}
  \label{fig:biasmass}
\end{figure}

To show why the bias of the unweighted halos agrees with the ST prediction while the luminosity-weighted biases
deviate from it, Fig.~\ref{fig:biasmass} gives the bias of the halos in mass bins of $0.1$\,dex at six snapshots,
measured from the cross-spectrum of each bin with the matter in the large-scale limit. At $z = 0.5$ the
measurement lies about a tenth below the ST prediction over the masses that carry the light; the offset shrinks with
redshift, changes sign near $z = 2.5$, and from $z = 3.5$ runs above the prediction, more so at higher mass.
The luminosity-weighted biases (Fig.~\ref{fig:biaslum}) inherit these offsets almost one for one.

We have not explained the rise of the measured bias above the prediction near the catalog floor, which carries the
agreement of the unweighted halo counts (Sec.~\ref{sec:bias}); a plausible cause is the halo finder, which links
groups of a few tens of particles more readily in dense regions and so would raise the measured bias of the
smallest halos above that of genuine halos of the same mass. The IHL samples the massive end instead, drawing half
its light from halos above $10^{13}\Msunh$ at $z = 0.5$ and above $7\times10^{11}\Msunh$ at $z = 5$: at high redshift it
samples $b(M)$ where ST falls shortest and its bias exceeds the prediction, while at $z = 0.5$ its light
straddles the mass where the offset changes sign and the two nearly cancel.

\section{The ingredients of the prediction}
\label{app:ingredients}

In this appendix, we quantify the impact of each modeling ingredient by replacing it with its naive alternative,
with everything else held fixed. The measured mass function, the box's clustering and the estimator's windows are
discussed in the body of the paper. Of what remains, four ingredients act on the IGL and the lines consistently
across the band, two properties of the construction act on the lines channel by channel, and the IHL raises issues of its own.

The measured bias changes the line auto-spectra by $10$--$25\%$ in their clustering-dominated bins, and the IGL's by $18\%$. Integrating through the redshift range of each shell, rather than evaluating it at its midpoint, changes the IGL's shot term by $25\%$; a line's channel is a slice much thinner than its shell, which needs no such integral. The LoS density of the stretched $z = 0.5$ slab changes the IGL mean by $5\%$ and its measured-to-predicted ratios by $4$--$10\%$, and affects a line only in the channels inside that shell (Sec.~\ref{sec:lim}). Holding the emission spectrum at its snapshot, as the maps do, changes the IGL auto-spectrum by $1$--$3\%$ but its fitted amplitude in the detectability statistic by $13\%$ (Sec.~\ref{sec:threemodels}); a line enters the prediction as a single luminosity per halo, which has no spectrum to sample. With every ingredient in place, the
measured mean intensity of the $2.2\um$ IGL channel is $0.992$ of the predicted one.

The first channel-by-channel property is the range of the emitters: the maps paint a line only for halos whose
H$\alpha$ falls in one of the 78 channels, so near the band edges the emitters of an interloper line fill
only part of a channel's redshift path. A prediction integrating the full path counts emitters not present in the
maps, and sits a factor of two high where the painted emitters cover half of it.

The second is a shell boundary inside a channel: nine to eleven of a line's 78 channels are painted from two
snapshots, none at $2.2\um$. The shells carry independent isometries, so the prediction adds the two
pieces without correlation (Sec.~\ref{sec:samecons}); in the cross-spectrum, an emitter correlates with $\kappa$ only
through its own shell's matter, which the Limber approximation cannot represent, so the prediction projects the cross-spectrum in the flat-sky
approximation~\cite{Gao2023} with the lensing kernel truncated to the shell. A boundary channel loses up to a quarter of
its cross-spectrum at the lowest multipoles and little at the highest; at $2.2\um$ the three lines lie far from any boundary and move by a few percent, while the $1.48\um$ channel of Sec.~\ref{sec:bands} straddles the boundary at $z = 1.25$
and without the split its auto-spectrum is high by a third to a half. The growth is likewise held at the
snapshot; a channel-averaged statistic is barely affected, but at $2.2\um$ the three lines lie on the near side of
their shells and the frozen growth lowers their cross-spectra by $4$--$9\%$.

The IHL needs a treatment of its own. The maps paint it as a point at each halo center, so its
one-halo auto-spectrum term is the shot noise of its weight, $f_{\rm IHL}L(M)$; a diffuse model, convolving the NFW
profile~\cite{Navarro1997} twice, would fall well below the maps at the highest multipoles. The prediction
therefore takes the two-halo term plus this point-source shot noise, whereas Paper~I~\cite{PaperI} models the
physical, diffuse IHL. For the IHL, the ratio that carries the box's clustering into the auto-spectrum (Sec.~\ref{sec:samecons}) has one caveat: over
$k \approx 1$--$2.5\hMpc$ for $z \le 4$ the IHL-weighted auto-spectrum falls up to $3\%$ below the Poisson shot noise of
its weights, so the ratio is negative there. The deficit is set by the size of the halos that carry the light,
not by the resolution of the mesh: it is the signature of halo exclusion, which
Ref.~\cite{MoradinezhadDizgah2022} models on large scales for line maps on these halos. We do not model the depth of the deficit.

\section{Pitfalls of the construction}
\label{app:defects}

In this appendix, we collect the defects of the construction that our comparisons exposed. The maps of the main text carry every correction.

\subsection*{Shells that share one box}

A lightcone cut from a single box must reuse its volume: two thirds of the pairs of slabs overlap, the
$z = 0.5$ slab contains all the others, and each part of the box appears in five shells on average. On large scales the density pattern grows without moving, so the shells are near-copies, and the Born sum of
Eq.~\eqref{eq:born} adds them coherently rather than in quadrature. Under the projection $\theta = x/\chi$ only a
structure near the center of the patch stays aligned between shells, so the correlation there is several times
what it is at the edges, reaching $r \approx 0.7$ for single pairs, and the extra power sits at low multipoles.

\begin{figure}[tb]
  \centering
  \includegraphics{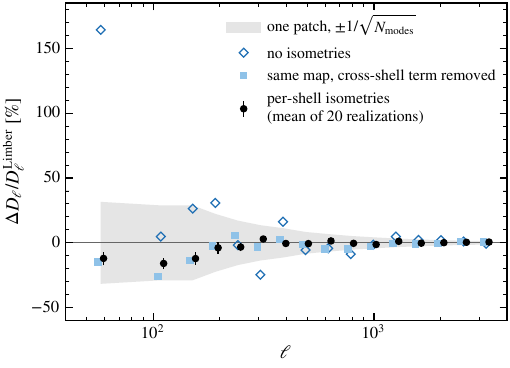}
  \caption{The fractional difference, in percent, of the convergence spectrum from its Limber prediction (halofit,
  $z < 6$). Open diamonds:
  one realization without isometries. Light-blue squares: the same map with the covariance between shells removed. Black
  points: the mean of twenty realizations with one random isometry per shell, with its error. Gray band: the single-realization Gaussian scatter. The three sets of points are offset slightly in $\ell$ for legibility. The realization without isometries places each slab at one
  distance, which changes its highest multipoles by $1$--$3\%$ relative to the cone projection.}
  \label{fig:shellcov}
\end{figure}

The power of the map splits exactly into the shells' own spectra, which is all a Limber prediction contains, and the covariance between them. Both follow from the maps of the individual shells, which makes the decomposition a test of the independent-shell assumption. Without isometries (Fig.~\ref{fig:shellcov}), the covariance carries two thirds of $C_\ell^{\kappa\kappa}$ in
the lowest bin and about a third near $\ell = 150$, and is consistent with zero above $\ell \approx 250$,
while the sum of the shells' own spectra agrees with the expectation of Sec.~\ref{sec:kappa} at every multipole, so that the independent-shell prediction fails and the map does not.

One random isometry of the box per shell (Sec.~\ref{sec:lightcone}) leaves the rms of the map and the shells' own
spectra unchanged and makes the covariance random in sign and consistent with zero in every bin; rotating the lightcone as a whole does not, since each slab must be rotated separately. In a lightcone without isometries, the shells cannot be treated as independent at low multipoles, and a jackknife over shells underestimates the error there.

\subsection*{Projecting the matter as a cone}

The halos of a shell sit at their own distances, so the matter of that shell must too. Placing the whole slab at
the snapshot's distance $\chi_{\rm cen}$ instead (Sec.~\ref{sec:lightcone}) shifts a structure at angle $\theta$
by $\Delta\theta = \theta\,|\chi/\chi_{\rm cen}-1|$, about $2^\circ$ at the edges of the $z = 0.5$ shell.
The matter map is then a slightly stretched copy of the sky the halos see, and the two stop correlating above
$\ell \approx 1/\Delta\theta$.

The cross-spectra take almost all of the damage: $C_\ell^{\kappa\kappa}$ moves by $1$--$3\%$ at the highest
multipoles, while in the two highest bins the IGL, the IHL and H$\alpha$ keep at most a few percent of their
cross-spectrum with $\kappa$. The nearest shells, whose relative
depth is largest, decorrelate first. Gaussian fields passed through the same stretch reproduce the measured loss with no
free parameter, and neither noise nor resolution can imitate its signature, a correlation higher at the center of the patch than in its corners.

The test is to follow one point through both projections. A single lit cell near the far edge of the $z = 0.5$
slab must land where a halo of the same box coordinate lands, and the correlation of each intensity map with its own
matter shell must not collapse at high \mbox{multipoles}. Two maps meant for cross-correlation should be projected in the same way, or a point source should be tracked through both projections. A shift changes no power, so a misregistration between the two maps is invisible to $C_\ell^{\kappa\kappa}$ and to every auto-spectrum, and the registration must be tested with a cross-spectrum.
\vspace{0.2in}

\section{The pixel scale}
\label{app:pixel}

In this appendix, we show that the $3'$ pixels of the maps (Sec.~\ref{sec:lim}) cost a cross-correlation with CMB
lensing almost nothing. The reconstruction noise of the SO baseline~\cite{Ade2019} exceeds the lensing signal
tenfold by $\ell = 1000$ and two hundredfold by $\ell = 3000$, so a finer map would have little to correlate with: in the forecast of Paper~I~\cite{PaperI}, stopping the sum over multipoles at $\ell = 3000$ instead of $5000$ lowers
the signal-to-noise ratio of the summed lines and of the IGL by at most $2\%$ and that of the IHL by $6\%$. The
simulated convergence also holds little structure \mbox{beyond} the pixel scale, since a $1\Mpch$ cell of the matter mesh
subtends $3'$ at $z \approx 0.4$.

The cost is that the maps test the predictions only up to $\ell \approx 3600$, and say nothing about the line
auto-spectra on the smaller scales where Ref.~\cite{Cagliari2025} draws most
of its constraining power, which is a question for line maps and not for a lensing cross-correlation. Little of the maps' own structure is lost. When the central degree of one realization is painted at SPHEREx's $6.2''$ pixel~\cite{Bock2025}, thirty times finer, it averages back to the $3'$ map to better than a percent in
the median IGL pixel, and its spectra are flat above $\ell \approx 9000$ for H$\alpha$ and $\ell \approx 2000$ for
the IGL, at the predicted shot noise.

\bibliography{refs}

\end{document}